\documentclass[fleqn,usenatbib]{mnras}
\usepackage{mathrsfs}
\usepackage{color}
\usepackage{mathptmx}
\usepackage[T1]{fontenc}
\usepackage{ae,aecompl}
\usepackage{mathrsfs}
\usepackage{svg}
\usepackage{bm}

\usepackage{graphicx}	
\usepackage{amsmath}	
\usepackage{amssymb}	
\usepackage{csquotes}   

\usepackage{tikz}
\usepackage{xcolor}

\definecolor{blueb}{HTML}{A1CAF1}
\definecolor{green}{HTML}{8DB600}
\definecolor{lightblue}{HTML}{87CEFA}
\definecolor{orange}{HTML}{FFA500}

\tikzset{conv/.style={black,draw=black,fill=blueb,rectangle,minimum width=6cm, minimum height=0.5cm}}
\tikzset{conv_block/.style={black,draw=black,fill=blueb,rectangle,minimum width=6cm, minimum height=0.5cm}}
\tikzset{conv_block2/.style={black,draw=black,fill=blueb,rectangle,minimum width=6cm, minimum height=0.6cm}}
\tikzset{conv_block3/.style={black,draw=black,fill=blueb,rectangle,minimum width=6cm, minimum height=0.8cm}}
\tikzset{conv_block4/.style={black,draw=black,fill=blueb,rectangle,minimum width=6cm, minimum height=1.0cm}}

\tikzset{pool/.style={black,draw=black,fill=green,rectangle,minimum width=4cm, minimum height=0.5cm}}
\tikzset{flatten/.style={black,draw=black,fill=lightblue,rectangle,minimum width=8cm, minimum height=0.5cm}}
\tikzset{bn/.style={black,draw=black,fill=green,rectangle,minimum width=6cm, minimum height=0.5cm}}
\tikzset{prelu/.style={black,draw=black,fill=lightblue,rectangle,minimum width=6cm, minimum height=0.5cm}}
\tikzset{dense/.style={black,draw=black,fill=orange,rectangle,minimum width=8cm, minimum height=0.5cm}}

\DeclareMathAlphabet{\pazocal}{OMS}{zplm}{m}{n}

\newcommand{\lya}{Ly$\rm{\alpha}$~} 
 
\newcommand{\kms}{${\rm km\,s}^{-1}$}
\newcommand{\mpc}{$h^{-1}{\rm cMpc}\,$}

\def\to{$\,T_{\rm 0}\,$}
\def\uo{$\,u_{\rm 0}\,$}

\newcommand{\zre}{$z_{\rm{re}}\,$}

\newcommand{\comment}[1]{}

\newcommand{\HI}{\hbox{H$\,\rm \scriptstyle I\ $}}

\newcommand{\HeII}{\hbox{He$\,\rm \scriptstyle II\ $}}

\newcommand{\logLag}{$-\log \mathcal{L}\,$}
\newcommand{\cov}{$\sigma_{\rm cov}\,$}

\newcommand{\optd}{${\rm log}{\Delta}_{\rm \tau}\,$}
\newcommand{\optt}{${\rm log}{\rm T}_{\rm \tau}\,$}

\newcommand{\den}{${\rm log}{\Delta}\,$}
\newcommand{\temp}{${\rm log}{\rm T}\,$}
\newcommand{\dto}{$\delta {\rm T_{\rm 0}}\,$}

\hypersetup{final}

\begin{document}
\title[Deep Learning the IGM]{Deep Learning the Intergalactic Medium using Lyman-alpha Forest at $ 4 \leq z \leq 5$}

\author[F. Nasir et al.]{Fahad Nasir$^{1}$\thanks{E-mail: nasir@mpia.de}, Prakash Gaikwad$^{1}$, Frederick B. Davies$^{1}$, James S. Bolton$^{2}$, Ewald Puchwein$^{3}$, \newauthor Sarah E. I. Bosman$^{1,4}$
\medskip
\\$^{1}$Max-Planck-Institut für Astronomie, Königstuhl 17, 69117 Heidelberg, Germany \\$^{2}$School of Physics and Astronomy, University of Nottingham, University Park, Nottingham, NG7 2RD, UK \\$^{3}$Leibniz-Institut für Astrophysik Potsdam, An der Sternwarte 16, D14482 Potsdam, Germany \\$^{4}$Institute for Theoretical Physics, Heidelberg University, Philosophenweg 12, D-69120, Heidelberg, Germany}

\label{firstpage}
\pagerange{\pageref{firstpage}--\pageref{lastpage}}
\maketitle

\begin{abstract}

Unveiling the thermal history of the intergalactic medium (IGM) at $4 \leq z \leq 5$ holds the potential to reveal early onset \HeII reionization or lingering thermal fluctuations from \HI reionization. 
We set out to reconstruct the IGM gas properties along simulated Lyman-alpha forest data on pixel-by-pixel basis, employing deep Bayesian neural networks. Our approach leverages the Sherwood-Relics simulation suite, consisting of diverse thermal histories, to generate mock spectra. 
Our convolutional and residual networks with likelihood metric predicts the \lya optical depth-weighted density or temperature for each pixel in the \lya forest skewer.
We find that our network can successfully reproduce IGM conditions with high fidelity across range of instrumental signal-to-noise. 
These predictions are subsequently translated into the temperature-density plane, facilitating the derivation of reliable constraints on thermal parameters.
This allows us to estimate temperature at mean cosmic density, \to with one sigma confidence \dto$\lesssim 1000{\rm K}$ using only one $20$\mpc sightline ($\Delta z\simeq 0.04$) with a typical reionization history. Existing studies utilize redshift pathlength comparable to $\Delta z\simeq 4$ for similar constraints.
We can also provide more stringent constraints on the slope ($1\sigma$ confidence interval $\delta {\rm \gamma} \lesssim 0.1$) of the IGM temperature-density relation as compared to other traditional approaches. 
We test the reconstruction on a single high signal-to-noise observed spectrum ($20$\mpc segment), and recover thermal parameters consistent with current measurements.
This machine learning approach has the potential to provide accurate yet robust measurements of IGM thermal history at the redshifts in question.

\end{abstract}

 

\section{Introduction} \label{sec:intro}

Accurately understanding the thermal history of the Intergalactic Medium (IGM) stands as a fundamental goal in astronomy, as it offers crucial insights into the timing and duration 
of the \HI and \HeII reionization epochs.
One of the most robust methods for probing this thermal history is through the examination of \lya absorption, commonly known as the \lya forest, observed in the spectra of quasars \citep{Becker_2015PASA}. This approach is grounded in the concept of photo-heating, where the IGM temperature increases due to \HI and \HeII ionizing photons \citep{Miralda-Escude1994}. The Doppler motions of the heated IGM are then imprinted onto the absorption lines of the \lya forest.

Leveraging on this idea, the literature has explored various statistics that demonstrate a range of sensitivity to the thermal state. Studies exploit \lya flux power spectrum suppression on small scales \citep{Zaldarriaga_2001ApJ,Croft_2002ApJ,Zaroubi_2006MNRAS,Viel_2013PRD,Walther_2019ApJ,Boera_2019ApJ}, the \lya line widths distributions \citep{Haehnelt_1998MNRAS, Schaye_2000MNRAS,Ricotti2000,Mcdonald_2001ApJ,Rudie_2012ApJ,Bolton_2012MNRAS,Bolton_2014MNRAS,Hiss2018,Telikova2019}, probability distribution of \lya flux \citep{Lidz_2006ApJ,Bolton2008, Calura2012,Lee_2015ApJ}, curvature of \lya flux \citep{Becker_2011MNRAS,Boera_2014MNRAS,Boera_2016MNRAS,Padmanabhan2015}, statistics of wavelet amplitudes \citep{Meiksin2000,Theuns_2000MNRAS,Zaldarriaga2002,Lidz_2010ApJ,Garzilli_2012MNRAS,Wolfson2021}, utilizing the entire $b-N_{\rm HI}$ distribution \citep{Hiss2019} and combining an ensemble of statistics \citep{Gaikwad_2021}. Nonetheless, despite recent progress, the pursuit of more precise methods persists, especially in the redshift range of $4\leq z\leq5$, which can provide insights into the early stages of \HeII reionization or any residual heating effects from \HI reionization.

The common theme of the previous studies has been to constraint the thermal parameters. Namely, the normalization (\to) and slope ($\gamma$) of the expected power-law relating temperature ($T$) and normalised cosmic overdensity, $\Delta$, of the IGM aftermath of reionization, $T=T_{\rm 0} \Delta^{\rm \gamma - 1}$ \citep{Hui_1997MNRAS,McQuinnSanderbeck2016}. 
While this approach has been valuable, it primarily offers a statistical description of IGM gas conditions. The richness of information present in the forest goes beyond what these thermal parameters can capture. Moreover, they are typically provided for the entire dataset at a given redshift. The methods are not sensitive enough to provide constraints for individual sightlines or to capture the variations within due to thermal fluctuations at $z \gtrsim 5$ \citep{D'Aloisio2015, Keating2018, Gaikwad_2020}. 
We aim to harness deep learning methods which can somewhat avoid the loss of information inherent to the standard summary statistics.

Recently, we have witnessed a notable surge in studies exploring deep learning (for review see \citep{Goodfellow2016}) in the context of quasar spectra. To highlight a few, generating quasar (missing) spectra based on its properties \citep{Eilers2022}, generating \lya forest for large surveys using only N-body simulation \citep{Harrington2022}, predicting \lya optical depth using transmitted flux \citep{Huang2021}, reconstructing the IGM temperatures on large scales around quasars \citep{Wang2022} and using \lya forest spectra to obtain constraints on IGM thermal parameters using ideal mock spectra at $z=2.2$ \citep{Parth2023arXiv}.

Our primary objective is to utilize deep learning techniques for the precise reconstruction of IGM densities and temperatures, pixel by pixel, within the \lya forest. By the aid of this method the thermal parameters can also be inferred from the reconstructed conditions. By virtue of sheer number of data points at our disposal, we can infer thermal parameters even for a single (20\mpc in this paper) sightline. Consequently, the constraints we obtain can be many times more powerful and accurate in comparison to traditional approaches. Moreover, since this approach strives to reconstruct the entire conditions along a given sightline, it naturally extends its capability to estimate thermal fluctuations that persist after the completion of \HI reionization or due to patchy \HeII reionization. 
However, we leave the analysis of inhomogeneous reionization models
to future work which would require an extensive suite of simulations for the training. We assume an IGM with a uniform thermal state for this paper.

There are few key considerations before we adapt deep learning methods for our purpose. We think it is crucial to incorporate uncertainties in particular for this reconstruction as actual conditions of IGM are unknown a priori. Quantifying the uncertainties due to lack of usable flux or due to parameter degeneracy becomes important, in particular at higher redshifts where the \lya forest is more opaque.
Furthermore, the optimal design of a neural network demands careful attention. A systematic grid search encompassing all relevant hyper-parameters allows us to harness the full potential of these techniques. Owing to their versatility and complexity, deep learning methods are susceptible to over-fitting issues. This implies that network performance may significantly deteriorate when applied to different but related datasets. Consequently, we take necessary precautions to mitigate over-fitting, such as rigorously testing performance on a dedicated test dataset using Nyx code.

In this work, we address and resolve the aforementioned challenges. We build a Bayesian Neural Network which maximises the likelihood of a given dataset with uncertainty estimates as a natural outcome.  We also mitigate the effect of over-fitting by carefully analyzing predictions on a test dataset. Furthermore, we conduct an extensive grid search to optimize the network architecture and all associated hyper-parameters for training.

Our paper is structured as follows: In Section 2, we provide details of the cosmological simulations employed in this study. Our focus is simulation outputs at redshifts $z=4-5$. Section 3 delves into the fundamental aspects of the neural network and finding an optimal network configuration. Section 4 present our core findings. We present the reconstructions along example \lya skewers. We also show recovery of thermal parameters through various models, using different signal-to-noise ratios. We also present a reconstruction of a 20\mpc segment on observational spectrum at $z=4$. In Section 5, we offer our concluding remarks.  We present a comparison of \lya optically depth weighted and real-space thermal parameters distributions in the Appendix.


\section{Hydrodynamics simulations}\label{sec:sim}
\begin{table}
  \centering
   \caption{The summary of simulations used in this work taken from Sherwood-Relics suite. The columns represent name, \HI photoheating rescaling factors, the redshift of reionization (the redshift where the global neutral fraction reach values below than $10^{-3}$), \zre. All simulations has box size length of 20\mpc with ${1024}^{\rm 3}$ gas and dark matter particles. The last row shows the runs which are used only for testing purposes.}
  \begin{tabular}{c|c|c}
    \hline
    \hline
    model & HI photoheating & \zre\\
    \hline
    fiducial/cold/hot & 1, 0.5, 2 & 6.0\\
    zr525/cold/hot    &1, 0.5, 2  & 5.4\\
    zr675/cold/hot    & 1, 0.5, 2 & 6.7\\
    zr750/cold/hot    & 1, 0.5, 2 & 7.4\\
    g10/g14/g16       & different $\gamma$ & 6.0\\
    \hline
    nyx-early/nyx-late & -- & 9.7, 6.66\\
    \hline
  \end{tabular}
  \label{tab:sims}
\end{table}

The neural network relies on appropriate datasets for training and validation purposes. We use a subset of cosmological simulations from the Sherwood-Relics \footnote{\url{https://www.nottingham.ac.uk/astronomy/sherwood-relics/}} suite, based on the Sherwood simulations \citep{Bolton_2017MNRAS}. 
These simulations were performed using a modified version of {\sc P-GADGET3} code an updated version of the {\sc P-GADGET2} \citep{Springel_2005MNRAS}. 
In this section, we will briefly describe the key features of these simulations which form our training and validation datasets. For details we refer the reader to \cite{Puchwein2023}.  These simulations are particularly designed for \lya forest absorption spectra with various reionization and thermal histories. In this work we restrict ourselves to runs with a standard Cold Dark Matter (CDM) cosmology. We summarise the runs we use in Table 1 (first five rows).

We use runs with a 20\mpc box length on each side evolved with $1024^3$ dark and gas particles. The simulations employ a spatially uniform but time-varying ultraviolet background (UVB) \citep{Puchwein2019},
and vary the redshift of reionization, \zre, and the thermal parameters, namely \to and $\gamma$. This is achieved by rescaling and shifting the photoionization and photoheating rates. There are 12 simulations on the \zre-\to grid (first row through the fourth row in Table~\ref{tab:sims}). Additionally, there are three runs for $\gamma=1.0, 1.4$ and $=1.6$ fixed at $z=4$ (fifth row in Table~\ref{tab:sims}). 
To estimate \to and $\gamma$ for a given simulation run, we calculate the median \temp on \den $=-0.4$ to $0.2$ bins with a size of $0.1$, and fit a line through these binned values. 
Our method is very similar to the one described in \cite{Gaikwad17}.

Finally, to verify the robustness and generalization of our network predictions, we include two more hydrodynamical simulations using the Nyx code \citep{Almgren2013} for testing purposes only. All runs have a box size of 20\mpc and were evolved with $1024^3$ gas and dark matter particles, using rescaled versions of the \cite{Haardt2012} UVB. 
The runs mimic very early and relatively late reionization histories as summarised in the last row of Table~\ref{tab:sims}.  The reionization and thermal histories of these runs do not closely match any of our training runs from Sherwood-Relics. We refer the reader to \cite{Onorbe2017} for more details.

We extract 20\mpc long skewers (in total $5000$) from each simulation running parallel to the box axes while tracing various quantities at $z=[4,\,4.4,\,5]$. We rescale the \lya optical depths to match the mean flux $\langle F \rangle=[0.4255,\,0.3216,\,0.135]$,  at each redshift which are taken from \lya forest measurements \citep{Becker_2013MNRAS, Bosman_2018MNRAS}. We convolve the spectra with Gaussian line profile with ${\rm FWHM}=6$\kms to mimic the resolution of VLT-UVES instrument. The pixel scale for our mock spectra is $\sim2.45$\kms at $z=4$. 
Note that, we add Gaussian distributed (or observational) noise with a fixed signal-to-noise during the training phase.

We form pairs of \lya flux and the corresponding logarithm of the \lya optical depth-weighted density and temperature (\optd, \optt) skewers for our dataset. It is infeasible to recover the real-space quantities (\den and \temp) from a velocity-space \lya flux skewer, as they are simply washed out by small-scale peculiar velocities of the absorbers due to structure formation. We use a Gaussian line profile (instead of a Voigt profile) to obtain \optd and \optt skewers. We found that in some rare sightlines, extreme high densities can affect the weighted quantity along the significant length of the skewer due to extended damping wing. The \lya forest is evaluated in the standard way using the Voigt line profile. We form one dataset by aggregating all skewers from the Sherwood suite (row one through five in Table~\ref{tab:sims}). 
The {\sc Nyx-early} and  {\sc Nyx-late} runs are only utilized at the testing stage. 


\section{Bayesian neural network}\label{sec:bcnn}
\subsection{Metric for network}


A network learns weights and biases during training using a gradient based optimization method. The main objective is that once the network is trained, the errors between actual and predicted quantities are minimized. This is solely judged on a metric and conventionally it is either mean absolute error or mean squared error.  However, as we want a handle on the uncertainties on predictions, we chose the negative logarithm of the Gaussian likelihood,  $-\log{\mathcal L}$ as our metric

\begin{equation}
    -\log{\mathcal L} = \frac{1}{N} \sum_{\rm i} \left( {(Y_{\rm i}-Y_{\rm i, pred.})}^{2} / \sigma_{\rm i,\, pred.}^2 + \log (\frac{1}{\sigma_{\rm i,\, pred.}^2}) \right),
    \label{eq:metric}
\end{equation}

\noindent here the sum runs over all the pixels. The metric is normalized by the total skewers in a training or validation split, $N$. The $Y_{\rm i}$, $Y_{\rm i, pred.}$ and  $\sigma_{\rm i,\, pred.}^2$ are actual, predicted mean and predicted standard deviation of the desired quantity, respectively. 



The underlying assumption for our metric is that, the predicted quantity follows a Gaussian distribution at every pixel along the sightline. Furthermore, we treat each pixel as independent and ignore any covariances at the training phase. This is to make training feasible and avoid predicting large covariance matrices which quickly become impractical for training a neural network. We later tackle this problem during the prediction step.  

\subsection{Building deeper networks with building blocks}
\label{sec:networks}
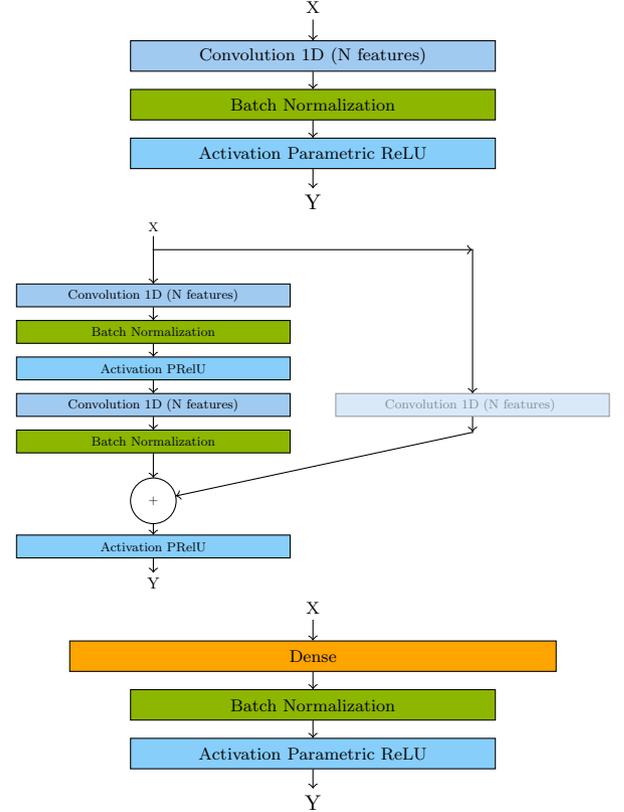
\begin{figure}
  \centering
\begin{tikzpicture}[scale=0.8, every node/.style={scale=0.8}]
    \node (x) at (0, 14.8) {X};
    \node[conv] (c1) at (0,14) {Convolution 1D (N features)}; 
    \node[bn] (bn) at (0,13.2) {Batch Normalization}; 
    \node[prelu] (act) at (0,12.4) {Activation Parametric ReLU}; 
    \node (y) at (0, 11.6) {\large{Y}};

    \draw[->] (x) -- (c1);\draw[->] (c1) -- (bn);
    \draw[->] (bn) -- (act);\draw[->] (act) -- (y);
  \end{tikzpicture}
  
\begin{tikzpicture}[scale=0.6, every node/.style={scale=0.6}]
    \node (x) at (0, 15.5) {X};
    \node[conv] (c1) at (0,14) {Convolution 1D (N features)};
    \node[bn] (b1) at (0,13.2) {Batch Normalization};
    \node[prelu] (p1) at (0,12.4) {Activation PRelU};
    
    \node[conv] (c2) at (0,11.6) {\text{Convolution 1D (N features)}};
    \node[bn] (b2) at (0,10.8) {\text{Batch Normalization}};

    \node[conv, opacity=0.4] (c4) at (7,11.6) {\text{Convolution 1D (N features) }};
    \node[draw,circle,minimum size=1cm,inner sep=0pt] (cir) at (0,9.5) {$+$};
    \node[prelu] (p2) at (0,8.5) {\text{Activation PRelU}};
    \node (y) at (0, 7.7) {\large{Y}};
    
    \node [draw,circle,minimum size=0.01cm,inner sep=0pt] (con1) at (0,15) {};
    \node [draw,circle,minimum size=0.01cm,inner sep=0pt] (con2) at (7,15) {};
    \node [draw,circle,minimum size=0.01cm,inner sep=0pt] (con3) at (7,11) {};
    
    \draw[->] (x) -- (c1);\draw[->] (c1) -- (b1);\draw[->] (b1) -- (p1);
    \draw[->] (p1) -- (c2); \draw[->] (c2) -- (b2); \draw[->] (b2) -- (cir);
    \draw[->] (cir) -- (p2);\draw[->] (p2) -- (y);
    \draw[->] (con1) -- (con2); \draw[->] (con2) -- (c4);
    \draw[->] (c4) -- (con3); \draw[->] (con3) -- (cir);
  \end{tikzpicture}

\begin{tikzpicture}[scale=0.8, every node/.style={scale=0.8}]
    \node (x) at (0, 14.8) {X};
    \node[dense] (c1) at (0,14) {Dense}; 
    \node[bn] (bn) at (0,13.2) {Batch Normalization}; 
    \node[prelu] (act) at (0,12.4) {Activation Parametric ReLU}; 
    \node (y) at (0, 11.6) {\large{Y}};

    \draw[->] (x) -- (c1);\draw[->] (c1) -- (bn);
    \draw[->] (bn) -- (act);\draw[->] (act) -- (y);
  \end{tikzpicture}
  
\caption[]{The schematic for the basic layers which forms the networks used in this work. {\it Top: } The Convolutional layer extracts N features using three pixels wide convolutional kernels. 
{\it Middle:} The Residual layer implements two stacked Convolutional layers with a skipping connection.  The input, $X$ is fed directly (or processed with convolution with same number of features if it is a first layer at that stage) to the last layer before activation. We stack several Convolutional or Residual layers to form one block, which placed in sequence to form either ConvNet or ResNet. 
{\it Bottom:}  The Dense layer performs dot product between the input and the weight matrix. All layers shares the batch normalization and Parametric Rectilinear Unit as activation which is an element wise operation. The Dense is the last layer in our network.}
\label{fig:building_blocks}
\end{figure}

\begin{figure}
  \centering
  \begin{tikzpicture}[scale=0.8, every node/.style={scale=0.8}]
    \node (x) at (0, 15.5) {\large{Normalized \lya Flux}};
    \node (s1) at (-5, 15.5) {\small{($N_{\rm batch}$x${\rm N}_{\rm sk}$)}};
    
    \node[conv_block] (c1) at (0,14.8) {\text{$N_{\rm 1}$ Convolutional/Residual Layers ($k_{\rm 1}$ features)}};
        
    \node (s2) at (-5, 14.4) {\small{($N_{\rm batch}$x${\rm N}_{\rm sk}/2$ x $k_{\rm 1}$)}};

    \node[conv_block] (c2) at (0,13.8) {\text{$N_{\rm 2}$ Convolutional/Residual Layers ($k_{\rm 2}$ features)}};

    \node (s3) at (-5, 13.4) {\small{($N_{\rm batch}$x${\rm N}_{\rm sk}/4$ x $k_{\rm 2}$)}};
    
    \node[conv_block] (cn) at (0,12.8) {\text{$N_{\rm n}$ Convolutional/Residual Layers ($k_{\rm n}$ features)}};

    \node (s4) at (-5, 12.4) {\small{($N_{\rm batch}$x${\rm N}_{\rm sk}/n$ x $k_{\rm n}$)}};

    \node[draw, shape = circle, fill = black, minimum size = 0.05cm, inner sep=0pt] at (0, 12.2) {};
    \node[draw, shape = circle, fill = black, minimum size = 0.05cm, inner sep=0pt] at (0, 12) {};
    \node[draw, shape = circle, fill = black, minimum size = 0.05cm, inner sep=0pt] at (0, 11.8) {};
    
    \node[flatten] (f) at (0,11.4) {\text{Flatten}};
    \node[dense]  (d1) at (0,10.6) {\text{Dense Layer (${\rm N}_{\rm sk}$ x 2 nodes)}};

    \node (mean) at (-3, 9.6) {\small{$\bm \mu$}};
    \node (s1) at (-3, 9.2) {\small{($N_{\rm batch}$x${N}_{\rm sk}$x2)}};
    \node (sigma) at (3, 9.6) {\small{$\bm \sigma$}};
    \node (s1) at (3, 9.2) {\small{($N_{\rm batch}$x${N}_{\rm sk}$x2)}};

    \draw[->] (x) -- (c1);\draw[->] (c1) -- (c2);\draw[->] (c2) -- (cn);
    \draw[->] (f) -- (d1);
    \draw[->] (d1) -- (mean);\draw[->] (d1) -- (sigma);

  \end{tikzpicture}
  \caption[]{The general schematic of our ConvNet or ResNet. Each stage is a stack of a certain number, $N_{\rm i}$, of Convolutional or Residual layers with a fixed number of features, $k_{\rm i}$. Each is processed with maximum pooling of two pixels in case of ConvNet or halving spatial samples with convolution (with stride two) in case of ResNet. The exact number of stages, layers at each stage and number of features are determined using hyper-parameters tuning.
  Here $N_{\rm sk}$ is the number pixels in a skewer and $N_{\rm batch}$ is number of skewers taken for each batch referred to as batch size. 
  The output format at each stage is shown on the left. The stage for each network is Dense layer.}
  \label{fig:schematic}
\end{figure}
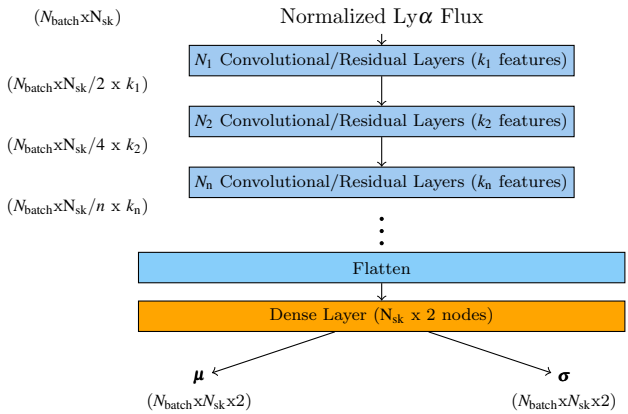

Before we tackle the problem of designing an optimal network, we will delve in to the basic building blocks for our networks. Three basic processing layers which are central to our networks are shown in Figure~\ref{fig:building_blocks}. We will refer to them as Convolutional (top), Residual (middle) and Dense (bottom) layers, respectively.

The main objective of the Convolutional layer is to extract a fixed number of features through a set of convolution kernels. We fix the kernel to be 3 pixels (we have also done trials with 5 and 7) typical for such networks. Not that, although the kernel size is set the subsequent convolutional followed with pooling aids the networks to learn complex features on larger scales.
The batch normalization layer standardizes samples of the current batch (a subsample of the training dataset) during training. For inhomogeneous datasets the network can be exposed to a biased batch, which can slow down the convergence of the metric during training. This layer helps to mitigate this effect. The activation layer introduces non-linearity into the network using the Parametric Rectified Linear Unit function (PRelu). This function retains input when it is positive but scales it with a trainable factor when negative. PRelu is a piece-wise linear function that has the advantage of being cost effective to compute. 

The underlying idea of the Residual layer is to skip connections between two consecutive convolutional layers by directly adding input to output before activation.
Note that, the inputs cannot simply be added when both can have a different number of features. Therefore, we process it with a simple convolutional layer (shown as a faded blue rectangle) to extract the same number of features (halving spatial samples with stride two if it is the first layer at a given stage) before adding it to the output.
This strategy of skipping layers has two advantages. If adding more layers is useful for the network, the gradients (calculated for error back-propagation) would be non-vanishing in deeper layers. If the additional layer has a negligible effect, the network can simply skip over them which saves time during training and does not affect network performance.

The first layer of the Dense layer implements a matrix multiplication between the weights matrix (learned during training) and the input. The Dense layer has the same two last layers as the Convolutional layer. The size of our Dense layer is twice the number of pixels in a \lya skewer, as the network predicts both the mean and standard deviation for each pixel. The basic difference between the Convolutional/Residual and Dense processing layers is the scale of features impacting the immediate output. The former extracts localized features while the latter takes advantage of full connectivity between inputs. 

We show the general schematic of our two networks, Convolutional Network (ConvNet) and Residual Network (ResNet) Figure~\ref{fig:schematic}. These networks are essentially 1 dimensional implementation of ResNet
\citep{He2015arXiv} using {\sc tensorflow/keras} \citep{chollet2015keras}.
The network performs processing in a total of $N$ stages. Each stage is a stack of 
$N{\rm i}$ Convolutional or Residual layers, where the number of features, $k_{\rm i}$, are kept fixed. 
The maximum pooling of two pixels is performed at the end of each stage in ConvNet. In ResNet, we scale down samples by performing convolution with stride 2, at the first convolutional layer at each stage. This scaling is shown on the left side of the schematic as down-sampling of each $N_{\rm sk}$, the number pixels in a skewer. 
We increase the total number of Convolutional/Residual layers and the features in each stage as we go deeper into the network. This is to tackle the increasingly complex features at later stages.  This progressive act of pooling/convolution and increasing features at each stage transforms data from sample to feature space. 
Both networks appropriately format (Flatten layer) the outputs and feed them to a Dense layer as a final processing stage. This layer outputs the mean and sigma of \optd or \optt for each pixel of a given skewer i.e $\bm{\mu}$ and $\bm{\sigma}$. 

\subsection{Training a Network}
\label{sec:train}

Now we focus our attention to details involving training a ConvNet or ResNet. In Section~\ref{sec:optimal_network}, we will discuss how hyper-parameters are tuned to find an optimal network architecture at $z=4$ through $5$.

We do a 80-20 percent split of the original dataset to separate the training and validation skewers. We have also done trials with 70-30 and 50-50 splits. All cases cases show very similar training histories but standard split show best convergence.
It is ensured that the both splits have proportionally the same number of skewers from each simulation run. 
Before training, both splits are standardized. This is achieved by subtracting the mean value and dividing it by the standard deviation of each quantity (i.e. \lya flux, \optd and \optt) of the training split. We use the same values to transform to or back when needed. This is a standard practice and helps to improve speed, stability and convergence during training. 

At the training/validation steps, we periodically shift the skewers to random locations taking into account the periodic boundary conditions of the simulation. This is equivalent to changing the starting position of a skewer along one axis. Later, we add Gaussian distributed noise with the fixed signal-to-noise pixel. The noise realization is generated during the training and validation steps. We do this procedure to overcome any over-fitting problems, which arise due to the limited number of skewers extracted along the box axes or because of a fixed noise realization.

The training and validation is performed in batches, the number of which is a hyper-parameter. Once the network has seen all the examples it is known as an epoch. The skewers in the entire dataset are randomized at the start of each epoch.
The metric is calculated first for training and later for validation split over all batches, with one value for each batch. The loss at each epoch is then given by a simple sum of the loss from all batches. This way we obtain training and validation losses for each epoch. We run training for 150 epochs which is enough to reach convergence of the metric. All training/validation sessions were ran on 4 A100 Nvidia GPU node with a typical completion time of under half an hour.  

\subsection{Finding optimal networks}
\label{sec:optimal_network}

The main hyperparameters we considered to search for our optimal networks are the type of architecture (ConvNet/ResNet), as well as the total number of
stages, layers and features at each stage. We impose a few restrictions in this grid search to avoid the network from becoming overly complex. We limit the total stages between 1 and 6. The number of layers at the first stage is chosen between 1 and 2.  It is then kept the same or doubled in subsequent stages, but never allowed to increase above 4. The number of features at the first block is chosen from $[2, 4, 8, 16, 32]$. It is either kept the same or doubled for each subsequent stage.

In addition, there are also hyperparameters related to the training which 
are needed to must be optimized. These are $N_{\rm batch}$ and the learning rate, $l_{\rm r}$.
The weights and biases of the network are updated at the end of every batch. A small batch typically results in noisy gradient estimation (calculated during error propagation), and therefore translates into fluctuations in parametric-space. A large batch gives an averaged gradient and the metric takes longer to reach optimal values. 
The gradient of weights and biases are adjusted through $l_{\rm r}$. A higher rate may introduce noise and stop the network from converging. On the other hand, smaller rates might make the network take way too long to reach convergence. We sampled $l_{\rm r}$ in log-space across $10^{-4}$-$0.5$. This is an initial learning rate however, we half its current value when the metric does not improve after 10 epochs. The batch size range in powers of two is $1024$-$8192$. The large batch size is chosen to run the training on 4 Nvidia A100 gups in parallel. Usually, there is a correlation between  $N_{\rm batch}$ and  $l_{\rm r}$, therefore, both parameters needed to be tune together.

We use {\sc Optuna}\footnote{\url{https://optuna.readthedocs.io/en/stable/}}, a python based API to search for optimal hyperparameters. We use the default Tree-structured Parzen Estimator provided by Optuna to sample the parameter-space.  We ran nearly 100 training and validation trails for \optt and \optd separately at each redshift. We are expecting each quantity to have a network with different complexity.  For each trail, we assemble a network with hyperparameters suggested by {\sc Optuna}. We train the network (procedure outlined in Section~\ref{sec:train}) and minimize our metric. We train for 100 epochs (50 epochs less than actual training) and keep the minimum value attained for \logLag for the validation split at the end of each trail. The code used for data preparation training and grid search is also available online at repository\footnote{\url{https://github.com/nicenustian/bh2igm}}.

We summarised the optimal set of hyperparameters of this extensive grid search in Table~\ref{tab:nn}. It is clear that a much smaller network extracting fewer features is preferable for \optd as compared to \optt. ResNet is the preferred network for all cases. Note that, the ResNet has twice the complexity as compared to ConvNet with same number of stages, layers and features. We have found that at higher redshift the network performance degrades which can be primarily attributed to low transmitted \lya flux. Although, there can be some impact on network architecture and (or) hyperparameters with the noise assumed for the mock spectra, we chose ${\rm S/N}=50$ per pixel as fiducial value.

\begin{figure*}\resizebox{17.5cm}{!}{\includegraphics[width=\columnwidth,trim={0.0cm 1.6cm .0cm 0},clip]{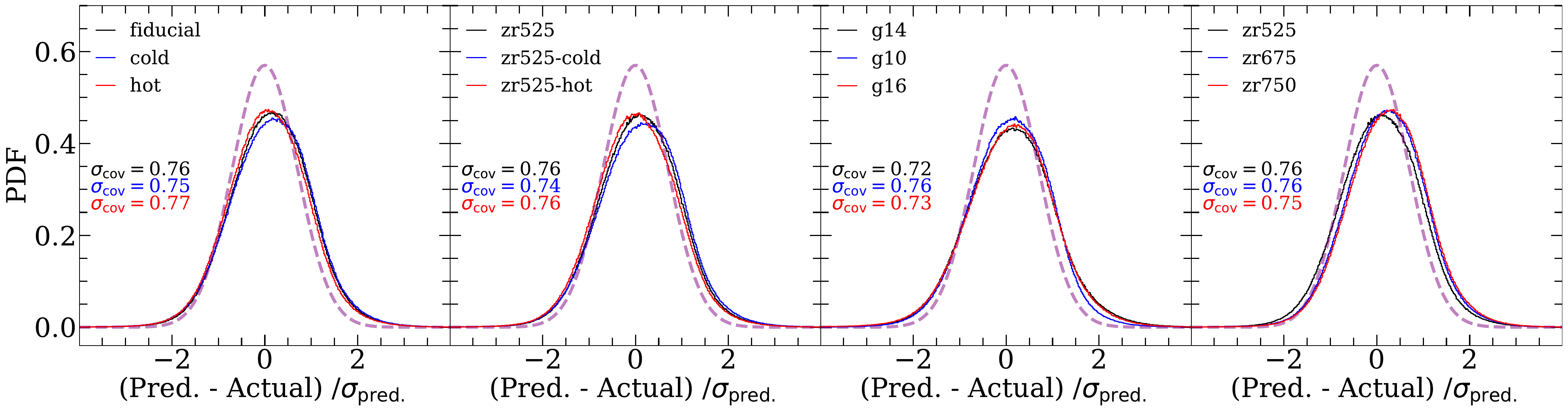}}
\resizebox{17.5cm}{!}{\includegraphics[width=\columnwidth,trim={0.0cm 0 .0cm 0},clip]{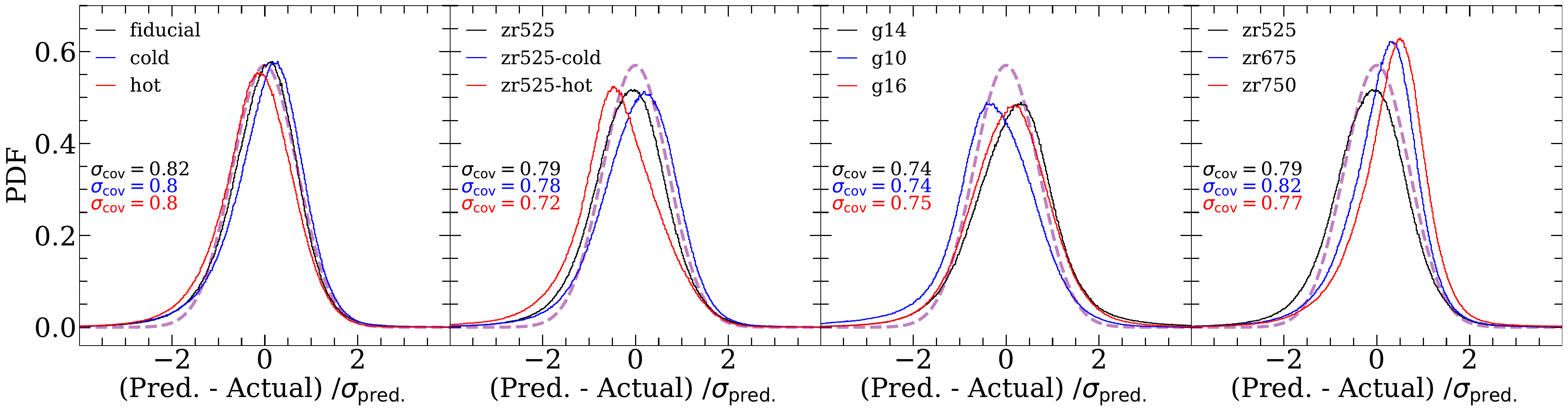}}
\vspace*{-2mm}
\caption{
The residual distributions of predicted \optd (top row) and \optt (bottom row) along with the respective \cov for the simulations shown in the legends. The dashed purple curve shows a Gaussian distribution with zero mean and unit variance to highlight any skewness or bias.}
\label{fig:residuals}
\end{figure*}

We have shown the residual distribution of \optd and \optt for each simulation in Figure~\ref{fig:residuals}. This plot highlights any bias and skewness of the predicted distributions among models with different thermal parameters. 

The \optd also shows a slight bias towards high densities, however for \optt it is most noticeable with different thermal histories. For instance, the distributions for {\sc hot} (red curve) and {\sc cold} (blue curve) are biased low and high respectively. The same is true for models with different redshift for reionization. {\sc g10} shows a significant tail, which can be attributed to relatively flat and insensitive \optt along the \lya flux skewer. Secondly, we are limited by training examples, as most of them are at $\gamma\simeq 1.2-1.5$. Although the predicted distributions for individual simulations can be biased, their covering fraction, \cov still remain above the expected $68$ percent.

\begin{table}
\caption{The hyper-parameters related to our networks used in this work at $z=4.0$ through $5.0$. The columns represents, layers at each stage, feature at each stage (progressively increasing in power of two), learning rate, $l_{\rm r}$ and batch size, $N_{\rm batch}$. All networks are ResNet in architecture.}  \centering
 \begin{tabular}{ccccc}
  \hline
  \hline
  $z$ & Layers & Features & $l_{\rm r}$ & $N_{\rm batch}$ \\\hline
  \multicolumn{5}{c}{\textbf{\optd}}\\\hline
  $4.0$ &  $1,4,4,4,4,4$   & $8,16,32,64,128,128$ & $7.1{\rm x}10^{-3}$ & $2048$ \\
  $4.4$ &  $2,4,4$      & $32,64,128$ & $20{\rm x}10^{-3}$ & $2048$ \\
  $5.0$ & $2,3,3,3$ & $32,64,64,128$ & $4.7{\rm x}10^{-3}$ & $2048$ \\
  \hline
  \multicolumn{5}{c}{\textbf{\optt}}\\\hline
  $4.0$ & $2,3,3, 4,4,4$  & $32,32,32, 32,32,64$ & $4.0{\rm x}10^{-3}$ & $2048$ \\
  $4.4$ & $1,2,2,  2,3,3$  & $32,64,128, 128,128,128$ & $2.2{\rm x}10^{-3}$ & $2048$ \\
  $5.0$ & $2,3,3, 4,4,4$  & $8,16,32, 32,64,128$ & $6.9{\rm x}10^{-3}$ & $8192$ \\
  \hline
  \end{tabular}
  \label{tab:nn}
\end{table}


\subsection{Predictions}
\label{sec:predictions}

For our results, we utilized predictions for all skewers that includes training and validation datasets from each model. The network outputs mean and standard deviations at each pixel of the \lya skewer for \optd and \optt.  However, using uncorrelated single pixel Gaussian's to reproduce realizations will give us unconstrained \to and $\gamma$ estimates. Therefore, we need to estimate a true error model which we could sample to measure \to and $\gamma$ and build their respective confidence intervals over many realizations.

\begin{figure}
\includegraphics[width=\columnwidth,trim={0.0cm 0.0cm 0.0cm 0},clip]{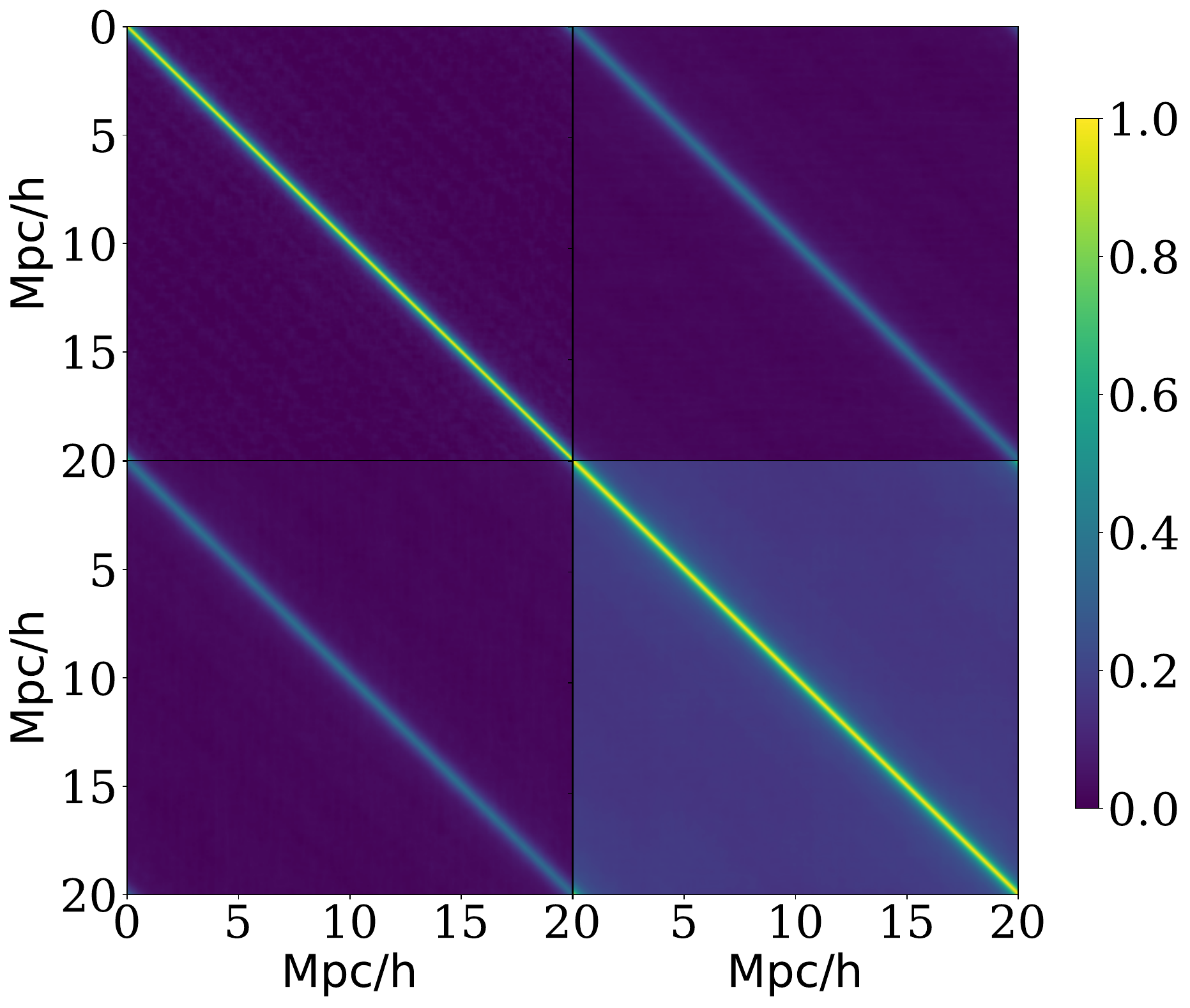}
\vspace*{-6mm}
\caption{The residual correlation matrix of concatenated \optd and \optt for {\sc fiducial} model at $z=4$. To determined a well behaved matrix 40 randomly shifted skewers from the model were stacked together. This amounts to $200,000$ skewers. The top-left represents \optd, while the bottom-right residual of \optt. The rest show the cross matrix between \optd and \optt. }
\label{fig:corr}
\end{figure}

We first obtain residuals of concatenated \optd and \optt, as $(Y_{\rm con}-{\bm \mu}_{\rm con})/{\bm \sigma}_{\rm con}$, where $Y_{\rm con}$ is the actual quantity, and ${\bm \mu}_{\rm con},\,{\bm \sigma}_{\rm con}$ are the predictions for the mean and standard deviation. 
Later, we congregate 40 sets of randomly shifted residuals to make the matrix smooth. This amounts to $200,000$ skewers in total. Finally, we obtain the residual correlation matrix, $\sum$. We repeat this process for each model and at each redshift. 
In Figure~\ref{fig:corr}, we have shown the correlation matrix for {\sc fiducial} model at $z=4$. It is evident that there is a significant correlation between residual pixels for \optt (bottom-right panel) at the box length scale. The \optt and \optd residuals are weakly cross correlated. Finally, we generate joint realizations of \optd-\optt skewers by simply sampling the multivariate Gaussian, ${\bm \mu_{\rm con}} + \mathcal{N}(0,\, \sum) {\bm \sigma_{\rm con}}$. We use the correlation matrix of the model with the least euclidean distance in \to-$\gamma$ plane.  However, we found no noticeable differences even when we obtain realizations using only the {\sc fiducial} correlation matrix. 
In practice, the confidence intervals obtained from a large number of realizations using this procedure and the ones directly from network predictions are extremely similar. However, individual realizations of a given skewer can differ significantly. We obtain $1000$ realizations for each skewer.

To estimate \to and $\gamma$ on \optd-\optt (for an actual or predicted realization), we make a small modification to our previous method. We employ an additional step of removing values which correspond to saturated pixels in the flux before fitting a line through the median \optt points in the desired \optd bins. We defined saturated pixels as when the flux is less than the 1$\sigma$ noise level. 
For each sightline we use all $1000$ \optd-\optt realizations and estimate \to-$\gamma$. This provides us with joint \to-$\gamma$ distributions for any given 20\mpc skewer.

\subsection{\lya Flux through the network}
\label{sec:cnn_flow}
\begin{figure*}
\resizebox{17.5cm}{!}{\includegraphics[trim={0cm 0cm 0cm 0cm},clip]{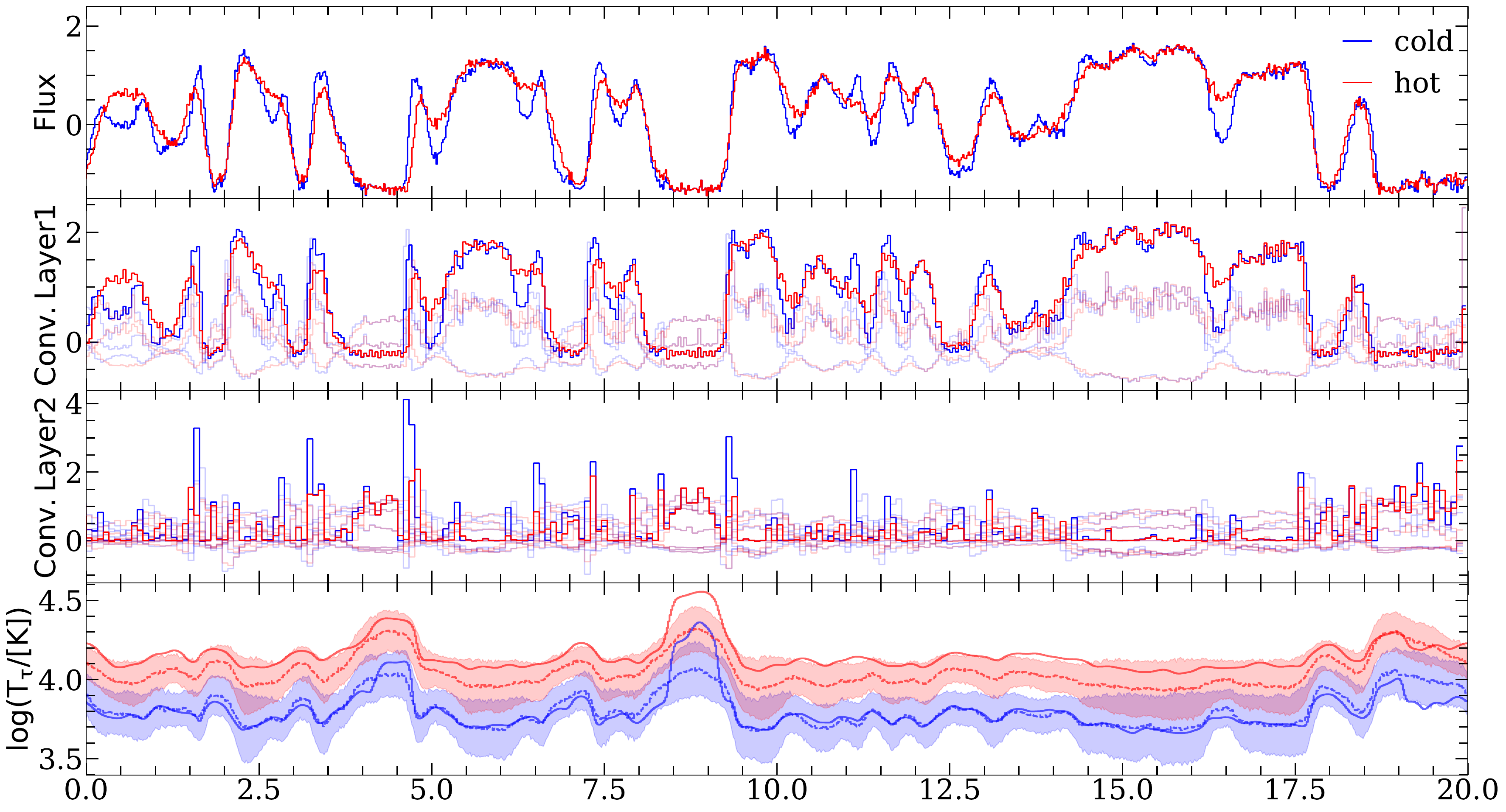}}
\vspace*{-2mm}
\caption{The flow of the example skewers at $z=4$ from 
{\sc hot} (red) and {\sc cold} (black) models through the various layers of trained ConvNet discussed in Section~\ref{sec:cnn_flow}. The first row is the normalized \lya forest skewer input to network. The second and third rows show the 4 and 8 features features respectively. The second Convolutional layer extracts each feature by combining all the skewers from first layer. We highlight the most prominent feature skewers for clarity. The output curves at the Dense layer (fourth row) shows predicted mean (dashed) and $1{\sigma}$ confidence intervals (colored regions), at each pixel of input skewers. The solid curves shows the actual \optt.}
\label{fig:cnn_flow}
\end{figure*}

In this section, we will examine the flow of the \lya flux through different layers of a simplistic network to build intuition into the reconstruction process.
In order to make the outputs easier to visualize at each stage, we utilize an elementary architecture. The network has two Convolutional layers, extracting 4 and 8 features respectively, forming a two stage ConvNet. We train the network using $N_{\rm batch}=32$ and $l_{\rm r}=10^{-4}$. We utilize the same dataset as our primary results for training and validation. 

Figure~\ref{fig:cnn_flow} shows the propagation of a normalised \lya forest skewer at $z=4$ through different stages. The purpose is to illustrate how the subtle differences in \lya flux between {\sc hot} (red curves) and {\sc cold} (black curves) models with different \to are translated into \optt predictions. 

The first convolution layer extracts 4 features directly from the normalised flux. The output is extracted by convolving the normalised flux using 3 pixel wide kernels which emphasizes the sharp features in the \lya flux of the {\sc cold} model.  Notice that the output pixels can be below zero which is only possible with PRelU activation. Allowing the neurons to fire even when the output is negative is crucial to fully utilize the dynamic range of normalised \lya flux pixels. 
At the second stage of convolution (third row), the trend is even more pronounced. Notice that each output skewer at the second stage is evaluated by combining all the feature skewers of the previous stage (in this case four) through one convolutional kernel. 

The fourth panel shows the output at the Dense layer. The feature skewers are finally transformed into predictions, which are distributions for each pixel. The network only outputs the parameters of the distributions which are the mean, ${\bm \mu}$ (dashed curves) and standard deviation, ${\bm \sigma}$. The predicted distributions are transformed back into their original units by using the mean and standard deviations from the training spilt.
We obtain 1$\sigma$ confidence intervals shown as light shaded regions.

The higher density regions show relatively less \lya transmission which translates into higher uncertainty in \optt or vice-versa. The actual \optt (solid line) falls mostly within the predicted confidence intervals (light grey region). It is expected that the actual quantity should fall within the predicted light-shaded contours at least $68$ percent of the time for the entire dataset. It is evident that the network sometimes fails to predict the right confidence intervals, specifically for saturated pixels. This give rise to the modest tails in the residual distributions.

\section{Results}\label{sec:fluc}

In this section, we will discuss in detail the predictions, primarily focused at $z=4$ and with ${\rm S/N}=50$ per pixel for noise (Section~\ref{sec:dist} to \ref{sec:thermal_params}). Our main goal is to recover the IGM \optd and \optt along the sightlines for our models with varying thermal parameters. 
This ultimately leads us to constrain the thermal parameters (i.e \to and $\gamma$). We also extend our analysis for spectra treated with different signal-to-noise (Section~\ref{sec:thermal_params}).  Later, we will extend the results up to redshift $z=5.0$ in Section~\ref{sec:thermal_params_redshift}. In the final section (Section~\ref{sec:obs}), we will show predictions for a segment taken from an observational spectrum and establish the method can provide reasonable but powerful constraints on thermal parameters.


\subsection{Predictions along the sightlines}\label{sec:los}

\begin{figure*}
\begin{center}
\resizebox{16.cm}{!}{\includegraphics[trim={0cm 1.5cm 0cm 0cm},clip]{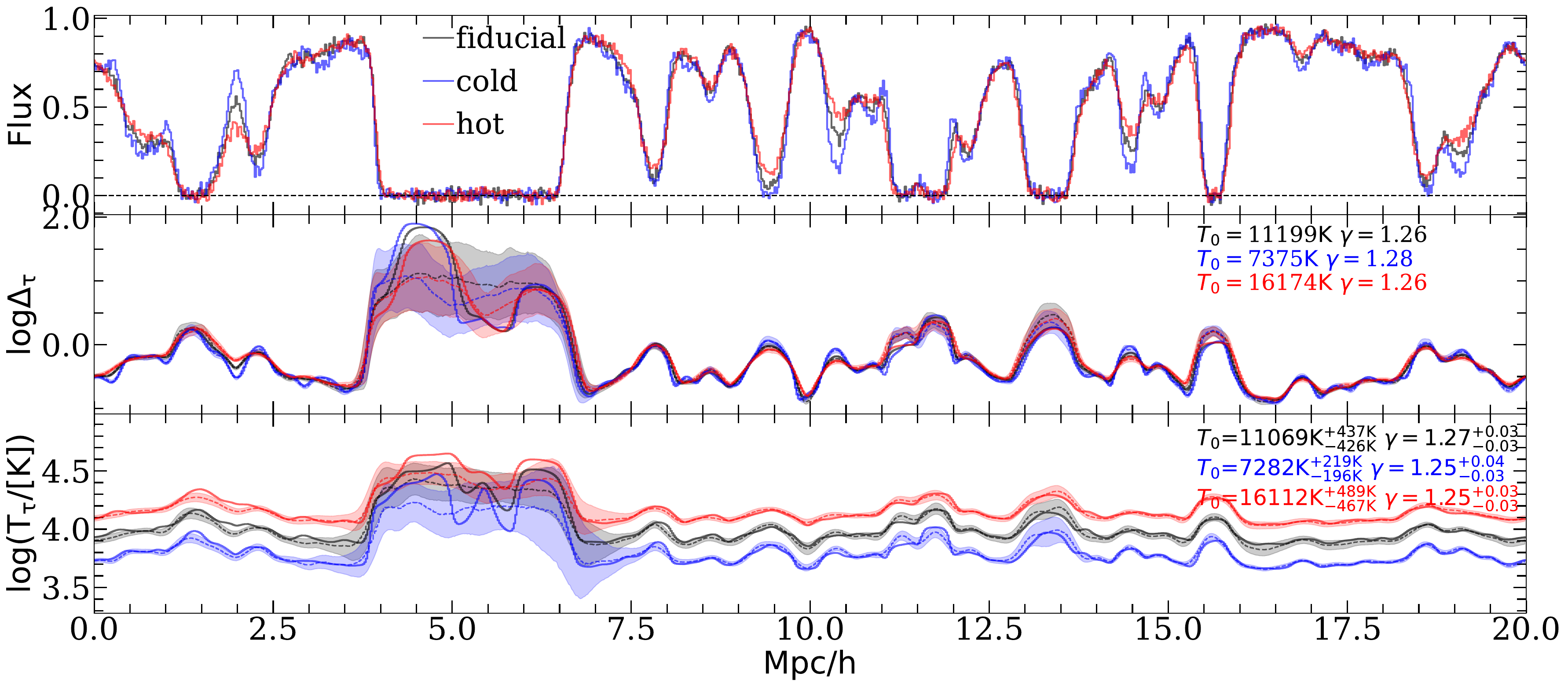}}
\resizebox{16.cm}{!}{\includegraphics[trim={0.0cm 0cm 0cm 0cm},clip]{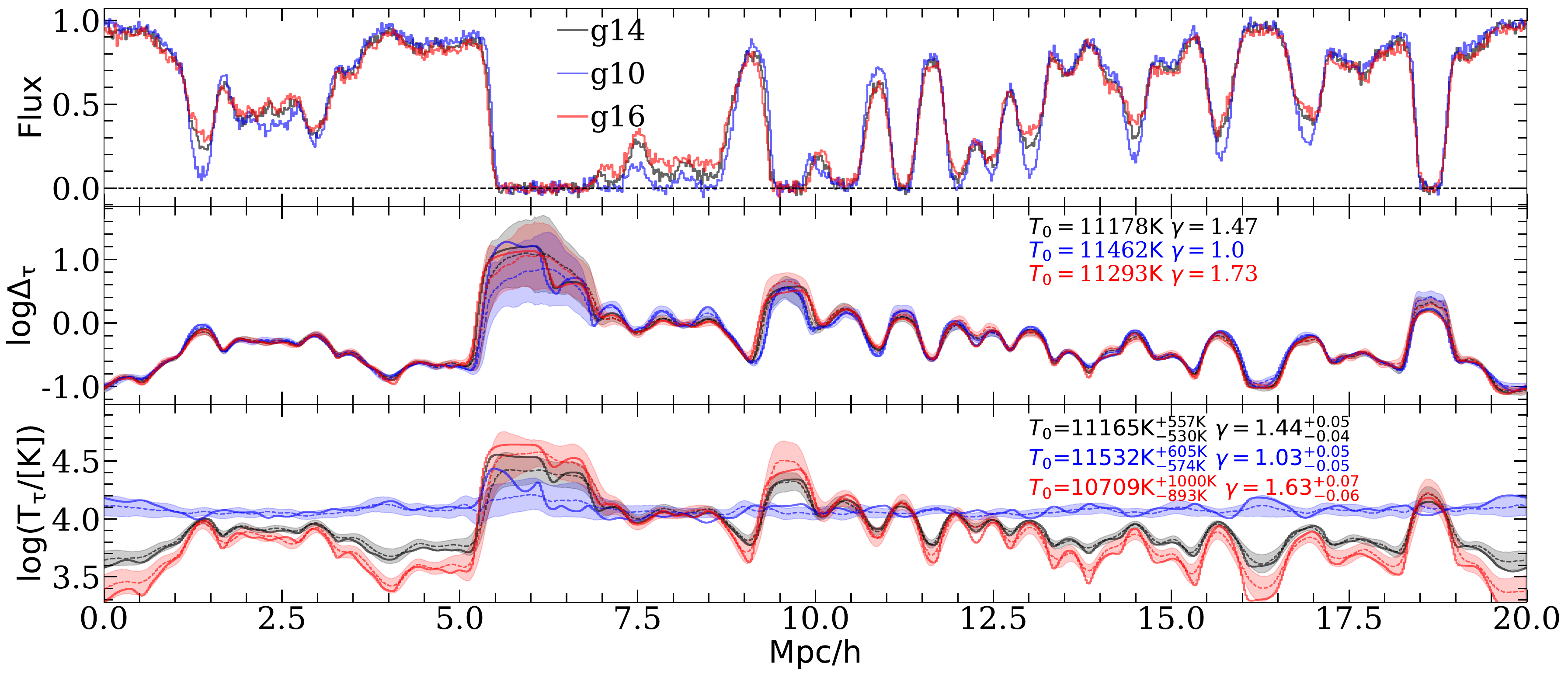}}
\end{center}
\vspace*{-4mm}
\caption{The predicted \optd and \optt along examples skewers for the simulations varying \to (top panel), $\gamma$ (bottom panel) at $z=4$ with ${\rm S/N}=50$ per pixel. Actual quantities are shown as solid curves overlaid with predicted mean (dashed curves) along with $1\sigma$ confidence intervals (colored bands). The actual \to and $\gamma$ is shown in second row of every panel, while estimated in every third row along with 1$\sigma$ confidence interval.}
\label{fig:skewers_t0gamma}
\end{figure*}

\begin{figure*}
\begin{center}
\resizebox{16.cm}{!}{\includegraphics[trim={0.0cm 1.5cm 0cm 0cm},clip]{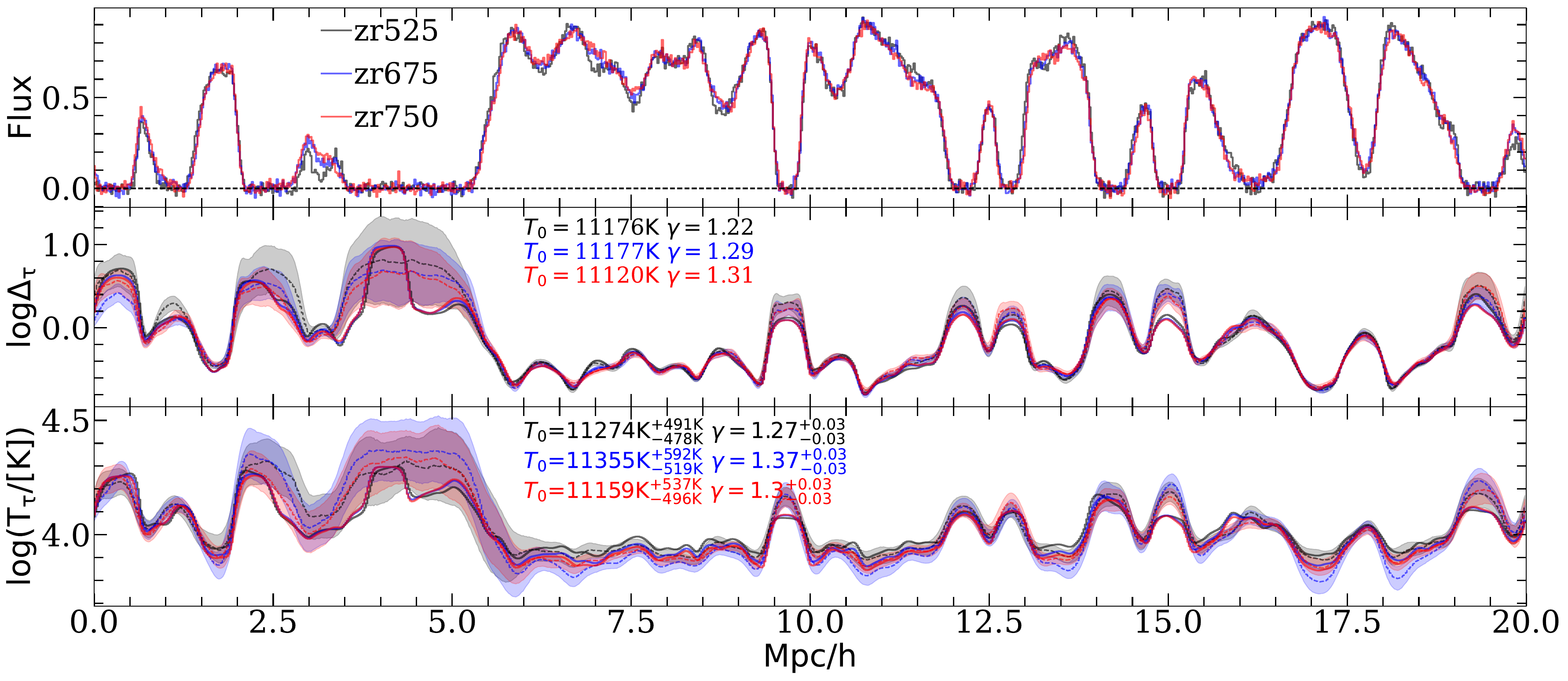}}
\resizebox{16.cm}{!}{\includegraphics[trim={0.0cm 0cm 0cm 0cm},clip]{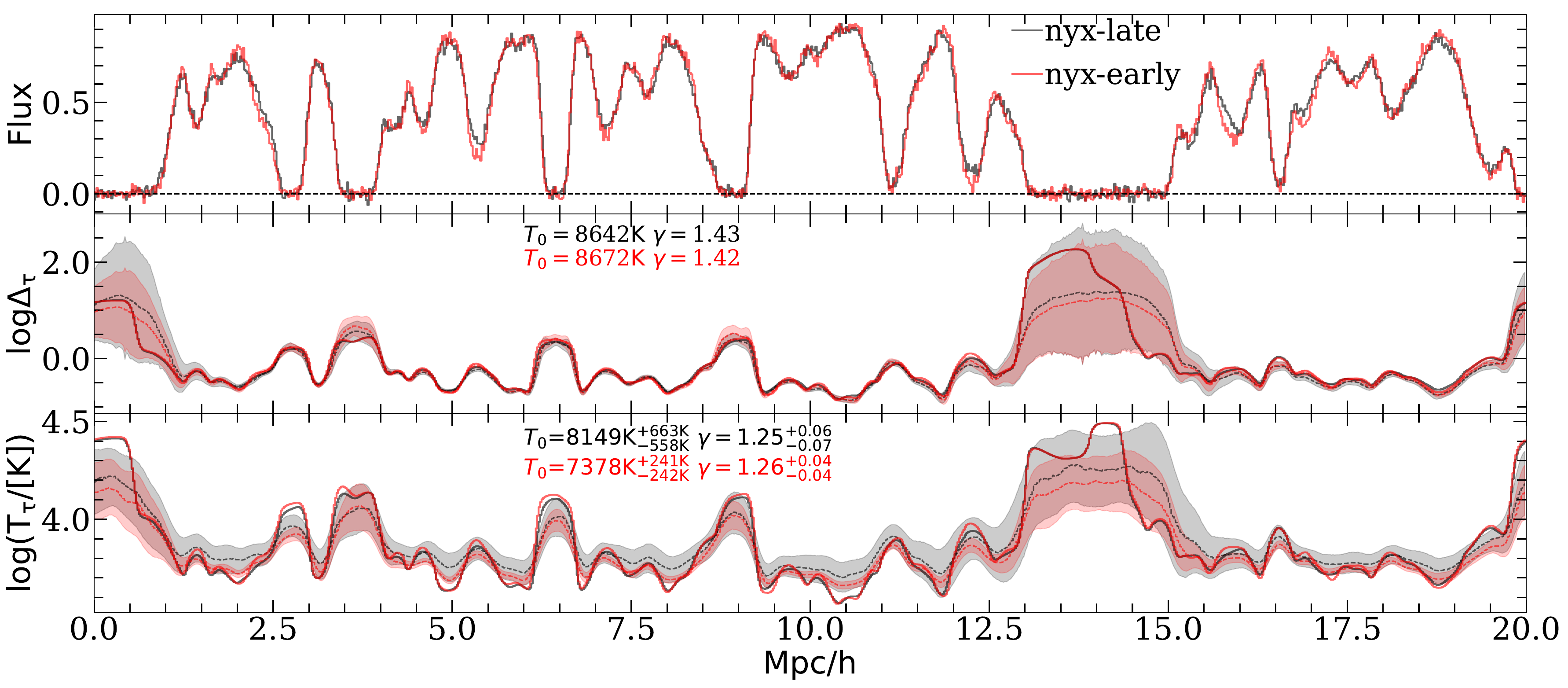}}
\end{center}
\vspace*{-4mm}
\caption{Same as Figure~\ref{fig:skewers_t0gamma} but now for runs varying \uo (top panel) and test models (bottom panel)}
\label{fig:skewers_u0nyx}
\end{figure*}

Figure~\ref{fig:skewers_t0gamma} and Figure~\ref{fig:skewers_u0nyx} shows the predictions of example skewers from simulations highlighting the impact of varying thermal parameters at $z=4$ with ${\rm S/N}=50$ per pixel. It is evident that the actual quantities (solid curves) lie mostly within the predicted 1$\sigma$ along the skewers. 

The 1$\sigma$ intervals are smaller in regions with significant \lya flux transmission and largest in the saturated parts. The network is unable to predict the right confidence intervals for saturated regions (for example at $5$\mpc in the first panel).  This shows the network is not over-fitting the training dataset. This essentially limits the predicting power primarily to under-dense IGM gas, which is not saturated at $z=4$.
The prediction for \optd has very narrow confidence intervals as compared to \optt. This is mainly because \optd reconstruction is very localised and is impacted by very local features extracted from \lya forest. The prediction of \optt at a given pixel depends on the \lya pixels on several scales. The small uncertainties on \optd predictions makes it potentially a method to constrain cosmological models which can impact the IGM densities on smaller-scales such as warm dark matter \citep{Irsic2017, Villasenor2023, Irsic2023}.

Varying \to impacts the small-scale IGM densities due to pressure smoothing ~\citep{Hui_1997MNRAS, Peeples_2010MNRAS, Nasir_2016MNRAS}. It is clear that the predictions also captures faithfully \optd along the sightlines (see Figure~\ref{fig:skewers_t0gamma} top-panel). The differences in \optd skewers are partly due to the difference in the \lya Doppler broadening between model varying \to.  
The {\sc cold} has noticeably more structure as compared to {\sc hot}. The impact is subtle but smaller uncertainties helps to reliably capture this in \optd predictions. The models with different \to also has slightly different slope of \to-$\gamma$ relation where {\sc cold} ({\sc hot}) is steeper(shallower) (see in Section~\ref{sec:thermal_params}). The predicted \optt along the very skewers remarkably trace the underlying temperatures with the right confidence intervals (except for saturated regions). 
The predicted \to values for the example skewers are predicted within a few hundred Kelvin of their actual values with 1$\sigma$ confidence intervals of \dto$\lesssim500{\rm K}$ for most cases.

The predicted \optt for models with varying $\gamma$ (second panel bottom row) can also captured the trend of actual \optt within predicted confidence intervals. The \lya flux  for {\sc g10} by and large insensitive to variations in \optt. However, the actual \optt still lies within narrow confidence intervals. The \zre parameter has a very fine imprint on the flux which simply translate in to larger uncertainties on \optt predictions. However, \optd is well constrained with similar uncertainties as compared to rest of the models. The subtle differences among these models is essentially due to Jeans smoothing in the gas.

Lastly, we have shown our test models {\sc nyx-late} and {\sc nyx-early} in the bottom panel of Figure~\ref{fig:skewers_u0nyx}. These runs were not included in our training or validation and do not impact the architecture or hyper-parameters. There is one point worth reiterating is that these models were run with entirely different hydro-dynamical code. The predictions are obtained with exactly the same method with frozen network weights. Qualitatively, the predictions are similar to those from the Sherwood runs. However, one difference is predicted \optt does not strongly correlates with \optd which results in systematically underestimated $\gamma$ value. We will later discuss this issue in Section~\ref{sec:temp_density_relation}.


\subsection{\optd and \optt distributions}\label{sec:dist}

\begin{figure*}
\begin{center}
\resizebox{17.5cm}{!}{\includegraphics[width=\columnwidth,trim={0cm 0cm 0cm 0.0cm},clip]{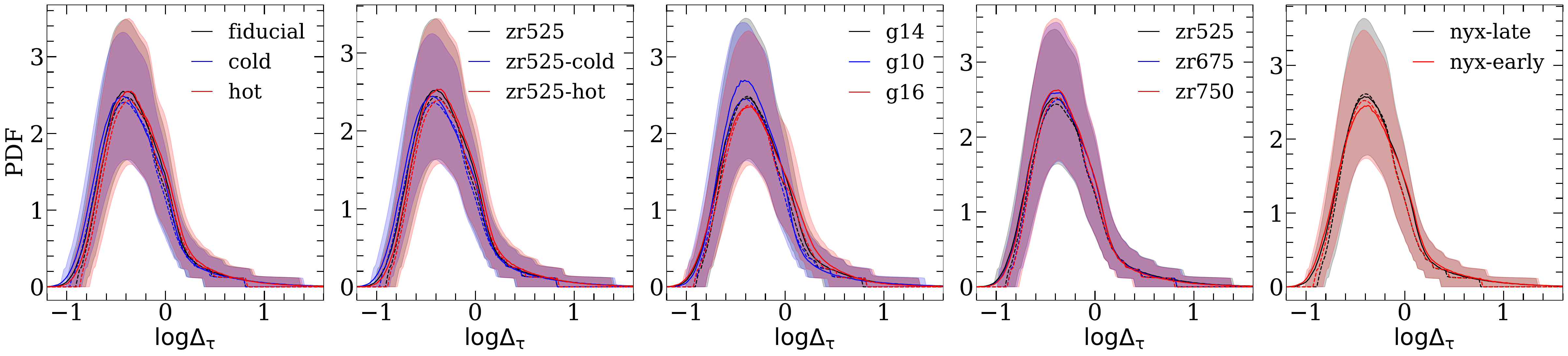}}
\resizebox{17.5cm}{!}{\includegraphics[width=\columnwidth,trim={0cm 0cm 0cm 0cm},clip]{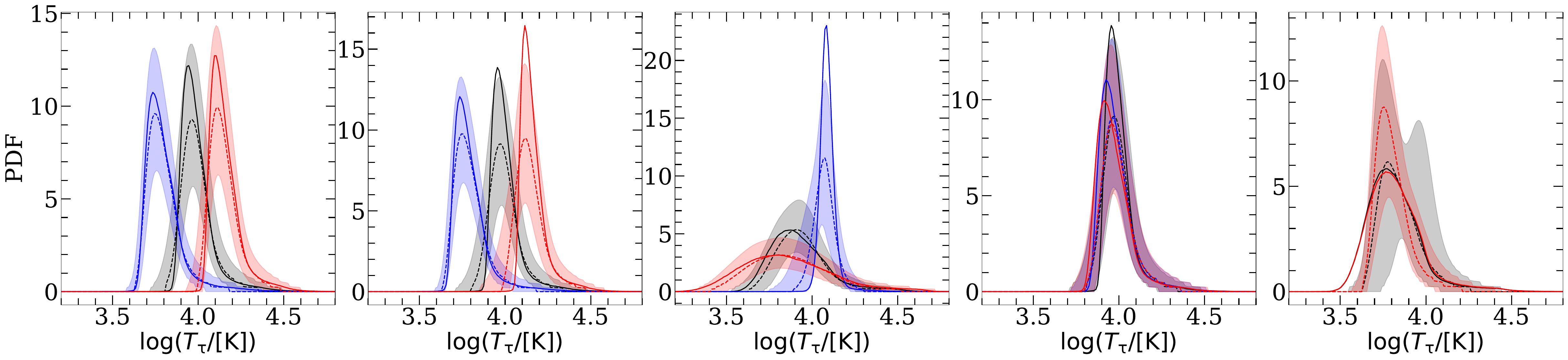}}
\end{center}
\vspace*{-4mm}
\caption{The distributions of \optd (top row) and \optt (bottom row) for 20\mpc skewers. The distributions for the entire skewers for actual and predictions are shown as solid and dashed curves respectively. The shaded regions represents the 1$\sigma$ scatter over realizations (see Section ~\ref{sec:predictions} for details). .}
\label{fig:dist}
\end{figure*}

To examine the range of IGM conditions probed by our reconstruction method, we examine the actual (solid curves) and predicted (dashed curves) probability density functions (PDF) in Figure~\ref{fig:dist}. we have overlaid the scatter in the distributions over 20\mpc skewers as the 1$\sigma$ shaded region in Figure~\ref{fig:dist}. Overall, there is a excellent agreement and no noticeable bias between the predicted and actual distributions. Despite some expected scatter among the 20\mpc skewers, the distributions can capture the broad trends in densities and temperatures probed by the \lya forest at $z=4$. It is worth mentioning that the step like feature at the high-\optd and high-\optt is  statistical, simply due to lack of pixels.

The \optd distributions are very similar for all models and are in agreement (see top row Figure~\ref{fig:dist}). The sharp drop in the distribution at mean cosmic density, \optd$\gtrsim0$ suggests that forest is mostly sensitive to under-dense gas at $z=4$.  The exact densities can slightly differ based on the simulation thermal parameters and history. This can be seen in models with different $\gamma$ (second last column) showing subtle differences at the high-density tail end. The actual distributions show a small excess at \optd$\simeq0.2$. The is primarily due to \lya flux saturation and consequently losing sensitivity at high-densities which ultimately degrades the reconstruction accuracy. The median of the predicted distributions ranges from \optd$=-0.33$ to $-0.37$, very closely following the actual range $=-0.32$ to $-0.38$.

Broadly, the predicted \optt distributions agree very well with a few exceptions. The {\sc g10} (third panel) is slightly narrow and exhibits a subtle offset. The {\sc zr525} is almost indiscernible from {\sc zr750}.
The {\sc nyx-late} model shows a bimodal distribution at the high \optt end.
As the shaded regions shows the variations you would expect from a 20\mpc sightline, it is evident that we can reliably estimate the IGM conditions with one sightline. We can expect a systematic bias which depends on the thermal parameters of model. However, this bias is typically small as compared to predicted confidence interval. Worth noting that the predicted \optt distributions are generally broader than the truth, but that this is expected given the predicted scatter.


\subsection{\optt-\optd plane}\label{sec:temp_density_relation}

\begin{figure*}
\resizebox{17.6cm}{!}
{\includegraphics[width=\columnwidth,trim={0cm 0cm 0cm 0},clip]{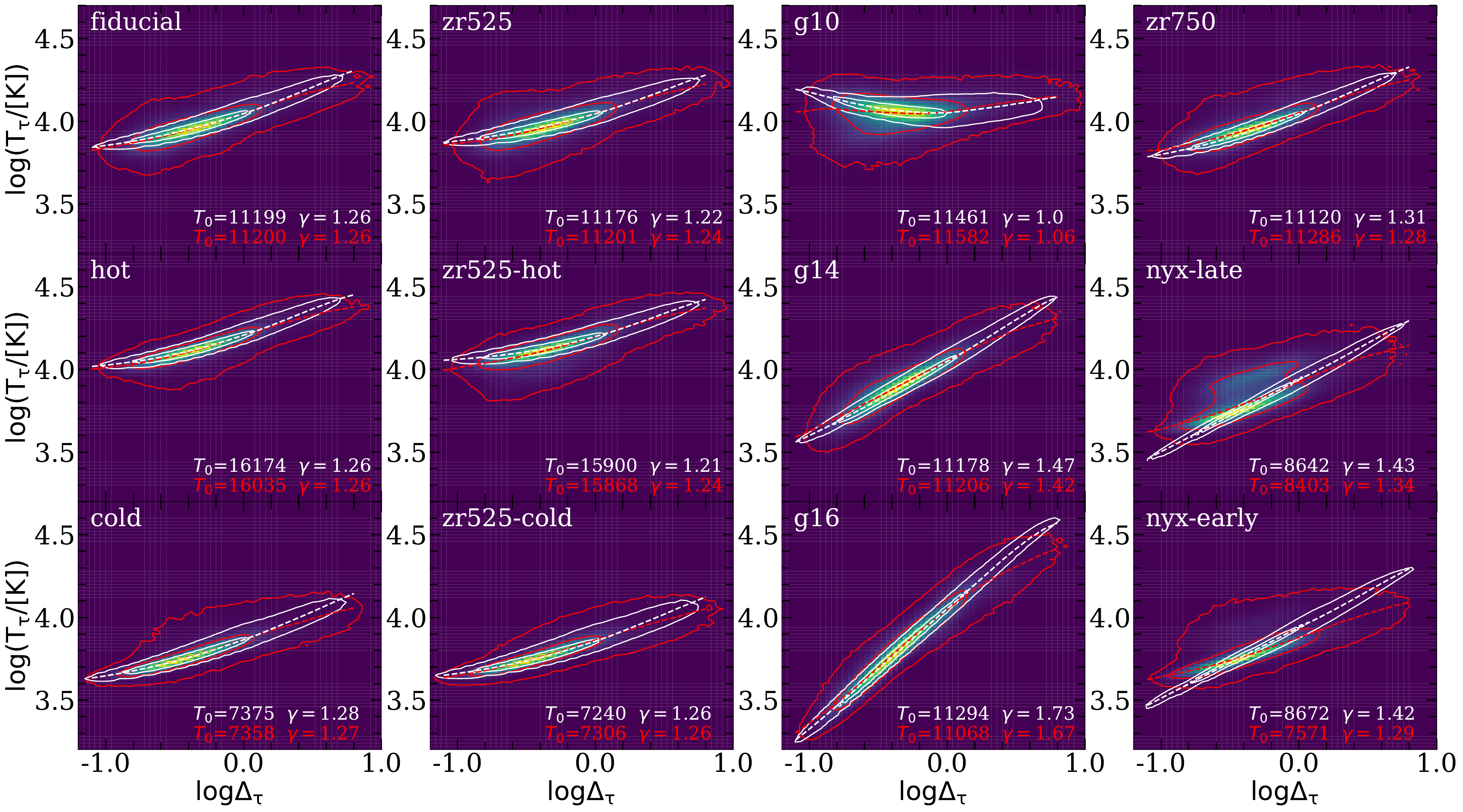}}
\vspace*{-2mm}
\caption{The predicted \optd-\optt distributions for the models shown in legends. The contours encapsulate the central $68$th and $95$th percentile interval for predicted (red contours) and actual (whites contours) quantities. The estimated \to and $\gamma$ for the entire skewers is also shown in legends. The dashed curves represents the median \optt on \optd bins. }
\label{fig:td_contours}
\end{figure*}

The characterization of IGM by thermal parameters (i.e \to and $\gamma$) although useful is an over-simplification. It is a fit to the complex 2-dimensional distribution of temperature and density. It is highly non-trivial to recover the entire \optd-\optt distribution and typically only thermal parameters are provided as a statistical insight. But owing to our reconstruction method, we can show the \optd-\optt plane and examine the detailed 2d distributions for models with varying thermal parameters. 

In Figure~\ref{fig:td_contours}, we have shown the predicted \optd-\optt distributions,  overlaid with $68$th and $95$th percentile for predicted (red) and actual (white) as contours. The median \optt on \optd bins (bin size $0.1$) are shown as dashed curves. The estimated \to and $\gamma$ are also shown and appropriately colored. By comparing the 2$\sigma$ contours, it is evident that the predicted distributions are more broader than the actual distributions . A typical \to is within few hundred Kelvins of the actual value, indicating that we can reliably estimate thermal parameters across various models. 

The median \optt shown as dashed curves broadly follows the trend but shows some noticeable deviations at low (\optd$\lesssim-0.4$) and high densities (\optd$\gtrsim0.4$). Therefore, in order to have a minimal biases on the thermal parameter estimates,  we fit a line through bins ranging from \optd$=-0.4$ to $0.2$ after removing the saturated pixels.
The distributions follow the trend we expect from the temperature-density relation in simulations evolved with non-equilibrium codes such as Sherwood-Relics (see \cite{Puchwein_2015MNRAS}). In general, we can reconstruct the deviations from simple power-law at lower-densities.

The test simulations ({\sc nyx-late} and {\sc nyx-early}) shown in last column exhibit a very narrow and rather steeper distribution. Recall, these simulations were evolved assuming photoionization equilibrium and therefore do not show any deviations from power-law at lower densities. Our predictions fail to capture these differences at the lower-density-end as highlighted by the median \optt curves shown in red. Another reason for this deviations is that we do not have any simulation run which is relatively steep $\gamma\simeq1.4$ but colder \to$\simeq8800{\rm K}$. Therefore, predictions show a rather lower $\gamma$ similar to our colder models such as {\sc cold} and {\sc zr525-cold}. These results can be improved with a more comprehensive model grid sampling the \to-$\gamma$ space.

\subsection{\to-$\gamma$ distributions}\label{sec:thermal_params}

\begin{figure*}
\resizebox{17.6cm}{!}
{\includegraphics[width=\columnwidth,trim={0cm 1.8cm 0cm 0},clip]{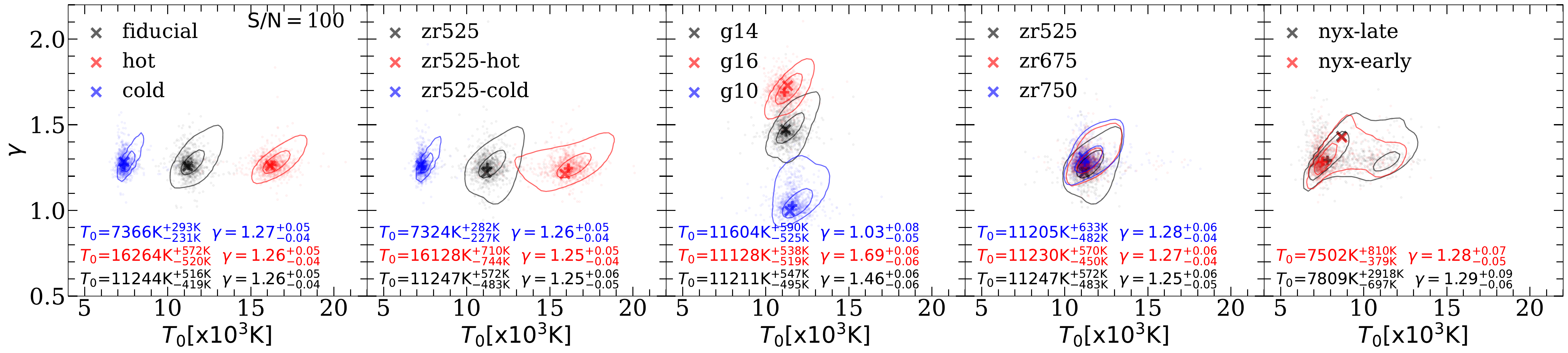}}
\resizebox{17.6cm}{!}
{\includegraphics[width=\columnwidth,trim={0cm 1.8cm 0cm 0},clip]{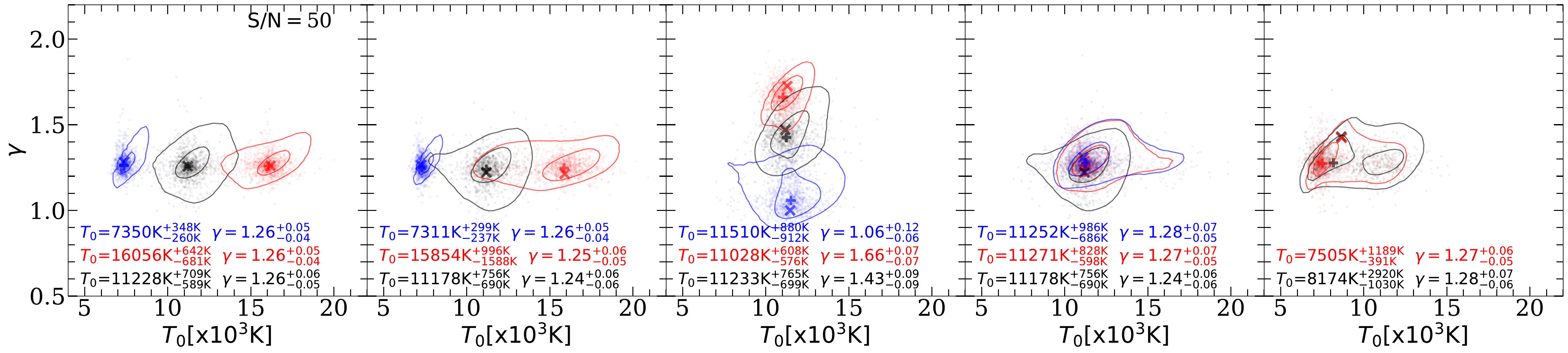}}
\resizebox{17.6cm}{!}
{\includegraphics[width=\columnwidth,trim={0cm .cm 0cm 0},clip]{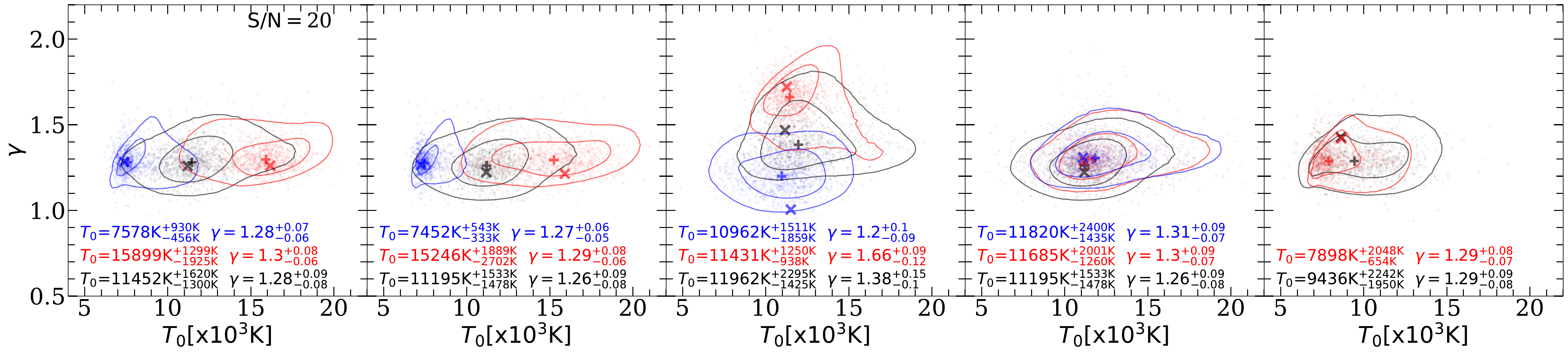}}
\vspace*{-2mm}
\caption{The \to-$\gamma$ distributions with different signal-to-noise ratios (shown in legends) at $z=4$. Each row represent sightlines which are post-processed with added Gaussian noise with ${\rm S/N}=100,\,50$ and $20$ per pixel respectively.
The distributions are obtained by estimating \to-$\gamma$ for each realization of 20\mpc sightlines ($1000$ realizations for each skewer). 
The contours encapsulates the central 68th and 95th percentiles of the points. The estimated \to and $\gamma$ along with their $1\sigma$ confidence intervals are shown in the legends. The plus and cross symbols mark the median of predicted and actual values for entire model.}
\label{fig:thermal_params}
\end{figure*}

\begin{figure*}
\resizebox{17.6cm}{!}
{\includegraphics[width=\columnwidth,trim={0cm 1.8cm 0cm 0},clip]{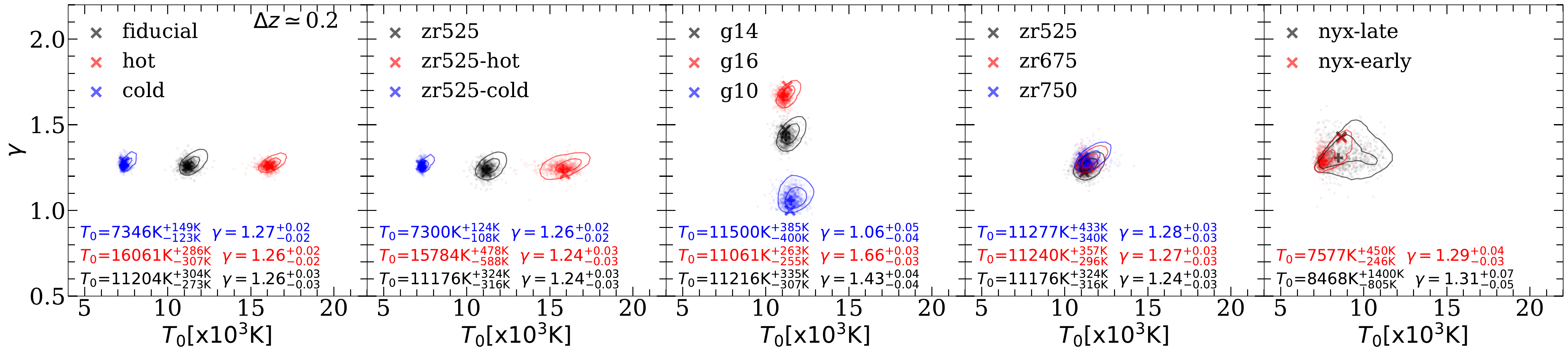}}
\resizebox{17.6cm}{!}
{\includegraphics[width=\columnwidth,trim={0cm .cm 0cm 0},clip]{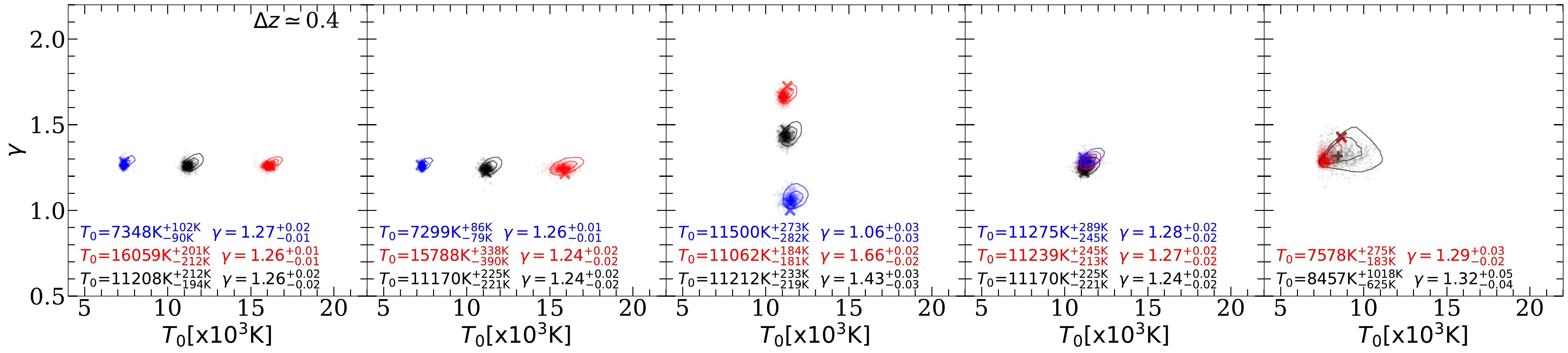}}
\vspace*{-2mm}
\caption{Same as Figure~\ref{fig:thermal_params} but now distributions are obtained by reiterating $100,000$ times over 5 (top) or 10 (bottom) \optd-\optt realizations with replacement from different skewers and estimating a joint \to-$\gamma$. This corresponds to total redshift path-lengths of $\Delta z\simeq0.2$ (top) and $\Delta z\simeq0.4$ (bottom). The signal-to-noise is kept fixed at ${\rm S/N}=50$ per pixel.}
\label{fig:thermal_params_dz}
\end{figure*}

We now proceed to show our main results of \to-$\gamma$ distributions for models. Recall, we have $1000$ \optd-\optt realizations for each 20\mpc skewer and their estimates for \to-$\gamma$. In total, we have $50,000$ \to-$\gamma$ data points for each model. We have shown the \to-$\gamma$ distributions (using all data points) using different ${\rm S/N}$ (first through third rows) in Figure~\ref{fig:thermal_params}. We add zero-centered Gaussian noise with a desired signal-to-noise during training/validation, as we have discussed in Section~\ref{sec:train}. The contours cover 68th and 95th percentiles of data points. A random selection of $1000$ data points is overlaid on the contours. The medians of actual and predicted values are shown as cross and plus symbol respectively. The predicted \to and $\gamma$ along with 1$\sigma$ confidence intervals for each model are shown in legends.

It is evident the predicted distributions are in good agreement with actual values (cross symbols) lying mostly within predicted 68th percentile.  The \to predictions are very constraining with typical uncertainties ranging between \dto$\simeq200{\rm K}-2300{\rm K}$. The $\gamma$ estimate also shows promise with $1\sigma$ uncertainties at $\delta{\rm \sigma}\simeq0.04-0.15$.

As expected, the constraining power degrades with decreased signal-to-noise. We can constraint our typical model with varying \to with $2\sigma$ confidence interval at ${\rm S/N}=100$ (except {\sc zr525-hot}), while at ${\rm S/N}=20$, it drops to $1\sigma$. The constraints for models varying $\gamma$ is less stringent.  For example, for ${\rm S/N}=100$ case the {\sc g10} is constrained with 2$\sigma$, while {\sc g14} and {\sc g16} with only 1$\sigma$ confidence. The {\sc nyx} models, are also borderline 1$\sigma$ contour. There is a systematic bias, with gamma under-predicted by $\sim0.1$. We have already discussed this issue in Section~\ref{sec:temp_density_relation}.

The uncertainties on thermal parameters also depend on the model.
By comparing models for ${\rm S/N}=50$ shown in first row, {\sc hot} has uncertainties at \dto$\sim1500{\rm K}$, while {\sc cold} is significantly lower \dto$\sim300{\rm K}$. Same is true for {\sc zr525-cold} and {zr525-hot}. Notice, the uncertainties are very similar for models with different $\gamma$ (similar \to) as shown in third column. The subtle differences in $\gamma$ for different \zre model can be seen in fourth column. Relatively late reionization {\sc zr525} tend to have small values for $\gamma$ as compared to early model {\sc zr750}, mostly due to IGM adiabatic cooling. The estimated median $\gamma$ hint about this evolution, although the uncertainties remain quite large.

So far, we have discussed the recovery of \to-$\gamma$ in the context of individual skewers. We can significantly improve these constraints if we consider combining several 20\mpc segments. For this, we combine their \optd-\optt realizations first and later we estimate the \to-$\gamma$ by fitting a line to the binned \optd-\optt as before.
To obtain the \to-$\gamma$ distributions, we simply draw $100,000$ times over any 5 or 10 skewers with repetition and estimate \to-$\gamma$. The resulting distributions for both cases are shown in Figure~\ref{fig:thermal_params_dz}.
The redshift pathlength for 5 (10) skewers is $\Delta z\simeq 0.2 (0.4)$ with fixed ${\rm S/N=50}$ per pixel.

As expected, the constraints get tighter with additional skewers, by comparing the \to-$\gamma$ distributions shown in Figure~\ref{fig:thermal_params} middle row (${\rm S/N=50}$ case) and Figure~\ref{fig:thermal_params_dz}.
For example, for models varying \to (first panel) the uncertainties are now reduced up to fifty percent for $\Delta z\simeq 0.2$ case. They are further reduced by $\sim 25$ percent for $\Delta z\simeq 0.4$, which we expect with the increase in pathlength according to central limit theorem.

The models with varying $\gamma$ (third panel) are now broadly enclosed within 95th percentile, but now highlight an important bias. Specifically \to bias in case of {\sc g16} and $\gamma$ bias for {\sc g10}. There is also noticeable improvements for the models with varying \zre (fourth panel). The noticeable reduction in uncertainties for $\gamma$ (up to $\sim50$ percent) helps to constrain the subtle $\gamma$ evolution with \zre.

Overall, the \to-$\gamma$ constraints provided by our method are more powerful as compared to the \lya forest flux power spectrum. 
Using our Bayesian neural network approach, a \emph{single} high resolution 20\mpc segment of the \lya forest can provide constraints on IGM temperature with uncertainties \dto$\simeq1000{\rm K}$ with a typically thermal history.  
This method potentially provide sub \dto$\simeq500{\rm K}$ constraints with a redshift path of $\Delta z\simeq0.4$, which is ten times lower than existing studies. In addition, the thermal parameters recovery for models varying $\gamma$ is also very encouraging. Although, a slight bias should be accounted for, in case of extreme $\gamma$.  A single skewer reconstruct can give sub $\delta \gamma \sim0.1$, which is usually achieved for considerable sized datasets using flux power spectrum studies in literature.

\subsection{Redshift evolution}\label{sec:thermal_params_redshift}

\begin{figure*}
\resizebox{17.6cm}{!}
{\includegraphics[width=\columnwidth,trim={0cm 1.8cm 0cm 0},clip]{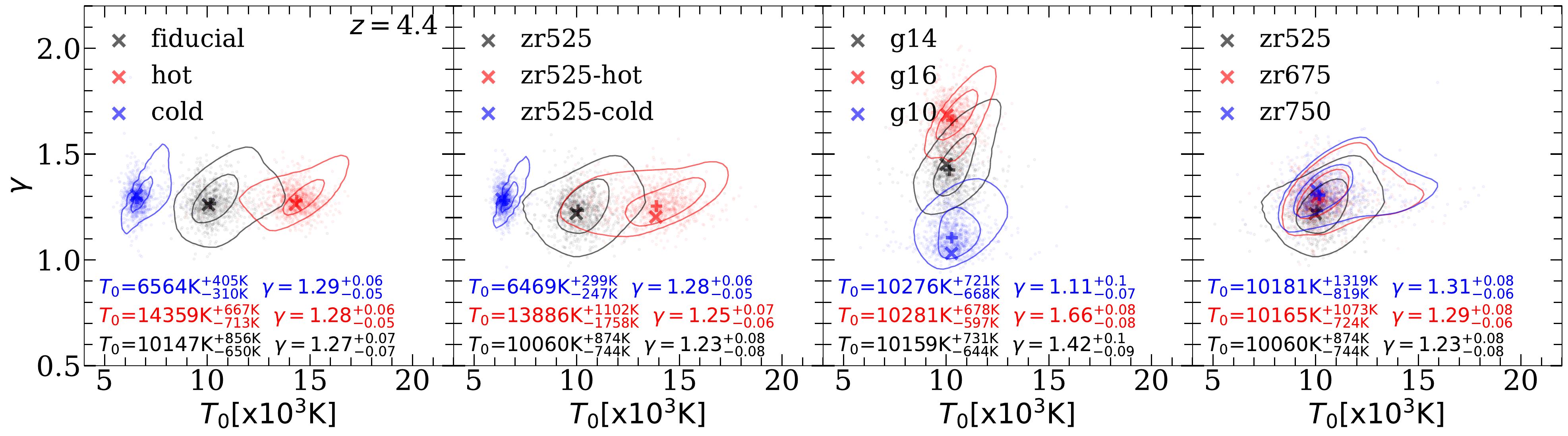}}
\resizebox{17.6cm}{!}
{\includegraphics[width=\columnwidth,trim={0cm 0.0cm 0cm 0},clip]{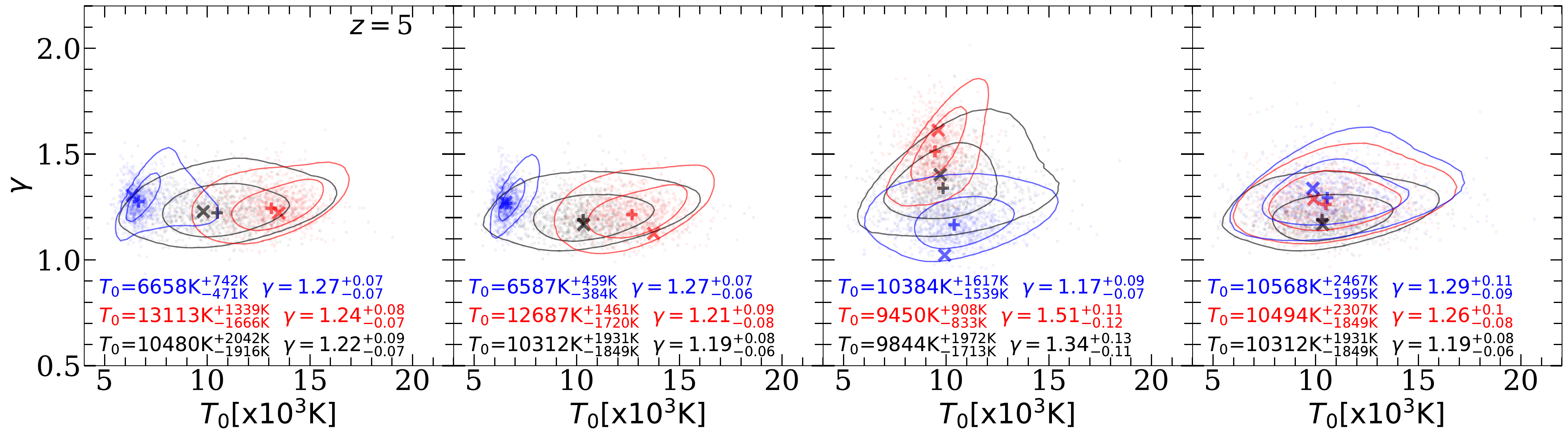}}
\vspace*{-2mm}
\caption{Same as Figure~\ref{fig:thermal_params} but at $z=4.4$ (top), and $z=5.0$ (bottom) with ${\rm S/N=50}$ per pixel.}
\label{fig:thermal_params_redshift}
\end{figure*}

Until now, we have have shown results at a fixed redshift $z=4.0$. To pose constraints on the thermal history of IGM, we need to extend this method to higher redshifts. As the \lya forest flux is the only input to our neural networks, it is likely that the performance could degrade due to significant drop in mean transmitted flux by $z=5.0$. In this section we will present our results at $z=4.4$ and $5.0$.

We show the \to-$\gamma$ distributions at $z=4.4$ (top row) and $z=5.0$ (bottom row) in Figure~\ref{fig:thermal_params_redshift}.
It is evident that the drop in mean flux impacts the constraints on \to and $\gamma$. Overall, there is a trend of rise in uncertainties with higher redshift. All distributions tend to become broader, which is most noticeable for models with varying $\gamma$ i.e {\sc g10} and {\sc g16}. 
By comparing the 95th percentile between redshifts, most of the distributions become broader in the \to direction. The increase in \dto can be up to $\sim50$ percent. Notice, due to the significantly broader distributions, models with different thermal parameters tend to overlap reducing the constraining power of the predictions using only single 20\mpc skewer.

\subsection{Observational sightline}\label{sec:obs}
\begin{figure*}
\resizebox{17.5cm}{!}{\includegraphics[trim={0.0cm .cm 0cm 0cm},clip]{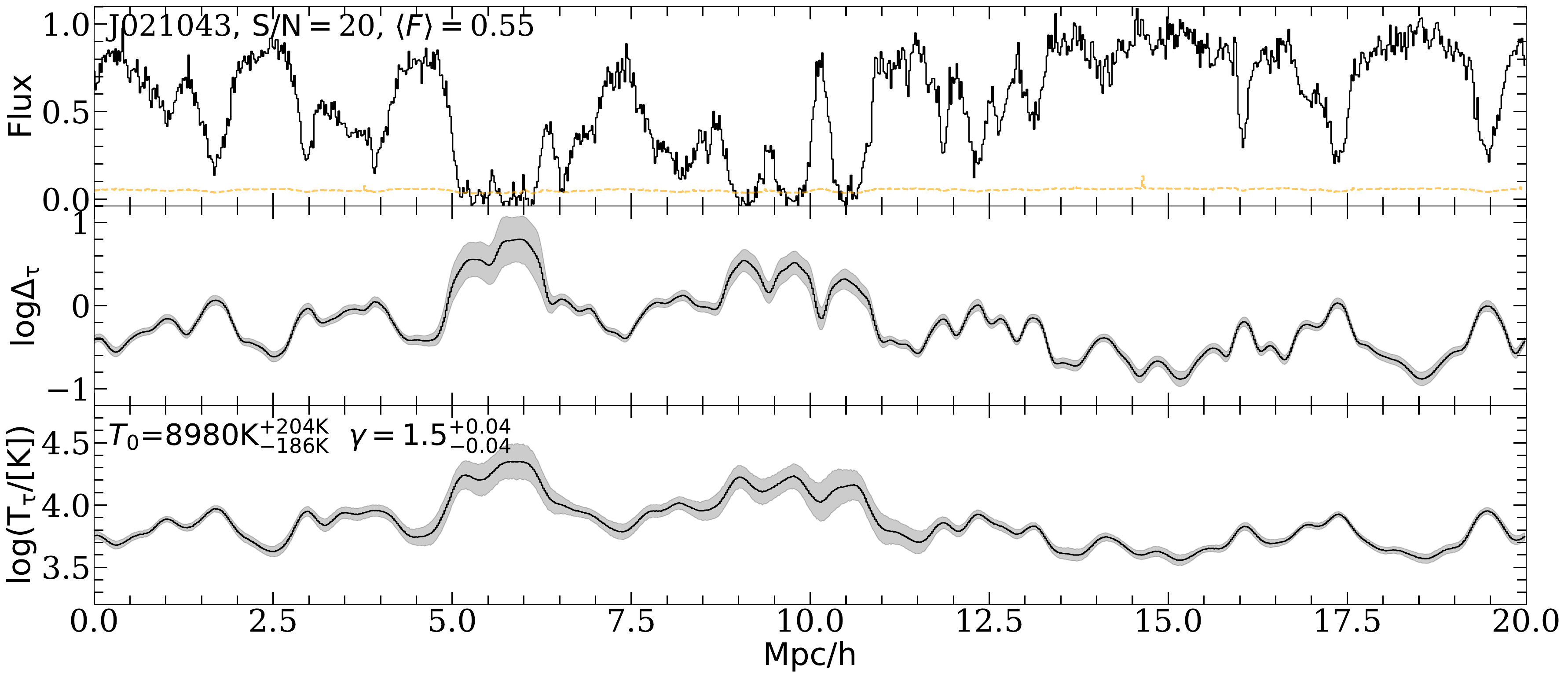}}
\vspace*{0 mm}
\caption{The predictions for IGM gas conditions of quasar J021043. This is 20\mpc segment (black) centered at $z=4$, overlaid with noise vector (orange). The predictions for \optd and \optt are shown in middle and bottom panels respectively. The predicted mean (dashed curves) along with $1\sigma$ confidence intervals (grey bands) are shown. The estimated \to and $\gamma$, with 1$\sigma$ confidence interval is also shown in the last row.}
\label{fig:los_obs}
\end{figure*}

So far, we have tested our method with mock spectra with different ${\rm S/N}$ and at different redshifts. In order to put method to practice and determined that any instrumental effects would not compromise our results
we take a 20\mpc segment from quasar J021043, which is part of {\sc squad dr1}\footnote{\url{https://github.com/MTMurphy77/UVES_SQUAD_DR1}} survey. The details of the reduction can found at \cite{Murphy2019}. The spectrum is observed using VLT-UVES instrument has resolution of ${\rm FWHM\sim}6$\kms with average  ${\rm S/N}=20$ per pixel over the skewer. The emission redshift of quasar is $z=4.65$. The spectrum is continua normalized and has bias regions removed for quasar proximity affect. 

To determine the appropriate noise realizations for the mock spectra of this observational spectra we determine a noise model by using the noise vector from sightline. As the noise is correlated with the transmitted flux level, we calculate median signal-to-noise in flux bins with bin size of $0.01$. Later, we add zero-centered Gaussian noise with the determined signal-to-noise based on the \lya flux of mock spectra during training/validation. We utilize {\sc fiducial} correlation matrix to estimate the \to and $\gamma$ with confidence intervals.

The predictions for \optd (middle) and \optt (bottom) for the \lya forest segment (top) overlaid with its noise vector (orange curve) is shown in Figure~\ref{fig:los_obs}. The mean (dashed curves) along with 1$\sigma$ confidence intervals as grey regions are shown.  The segment has a mean flux of $\langle F\rangle=0.55$ and ${\rm S/N}\simeq20$ per pixel. The estimates for the thermal parameters are \to$=8980{\rm K}{\rm}^{+204 \rm K}_{-186 \rm K}$ and $\gamma=1.5{\rm}^{+0.04}_{-0.04}$. The recovered value for \to is very similar to our {\sc cold} model. 
We want to reiterate the fact that only a single 20\mpc skewer realizations are used for estimating any \to-$\gamma$ distribution. One skewer corresponds to a redshift pathlength of $\Delta z\simeq0.04$ at $z=4$. For reference, the existing measurements at $4\leq z\leq 5$ utilized redshift pathlength of $\Delta z\simeq4$ or $6$, typically using $15$ or more high-resolution quasar spectra \citep{Boera_2019ApJ, Walther_2019ApJ}.
The measurements we have obtained from this example are also consistent with earlier measurements using summary statistics of the \lya forest \citep{Becker_2011MNRAS,Boera_2019ApJ, Walther2019}. However, we do not intend to present this result as a measurement but rather a way to validate the method. In future studies, we plan to apply this method to a more comprehensive quasar dataset at $z=4-5$ to measure the IGM thermal history in unprecedented detail.

\section{Conclusions}\label{sec:conclusions}

We have established that reconstruction of IGM gas conditions using Bayesian neural networks offers significant advantage over traditional summary statistics.  The method helps us to transform the \lya transmitted flux directly to (optical depth-weighted) gas densities and temperatures. This pixel by pixel reconstruction enable the mapping of the entire \optt-\optd plane. This is not possible with traditional methods which only recover thermal parameters using statistics such as the \lya flux power spectrum. We have shown that only one 20\mpc segment of the \lya forest from a single quasar can deliver constraints comparable to moderately sized datasets usually employed for such studies. We have seen that our method can provide fairly robust constraints on thermal parameters even in the presence of significant instrumental noise. We can also perform a reasonable reconstruction with test dataset from Nyx simulations by our trained neural network with frozen weights. In addition, the method can be extended up to redshift $z=5.0$, providing valuable insight into the thermal evolution of IGM. However, we expect the performance to get worse with mostly saturated spectra, therefore it might require a significant changes in the current architecture. The technique can also be pushed towards lower redshifts $z\lesssim4$, until most of the flux is at the continuum level.

Neural networks utilize quite complex feature-space transformation to convert \lya flux to the IGM conditions. This can potentially make inference somewhat more model dependent than traditional methods. For instance, traditional methods cannot capture the small deviation from a power-law in the temperature-density plane using the statistical thermal parameters \to and $\gamma$. 
As our method can reconstruct the entire plane, it requires the mock spectra to be a realistic representation of observations. This includes incorporating all the physical process which can impact the IGM conditions such as non-equilibrium photoionization. 
An important aspect of training large neural networks is to come up with clever solutions to over-fitting problems. We found that training can be sensitive to noise realization, skewers with correlated density structures and more importantly the network architecture. We have taken appropriate measures to over-come these problems by adopting strategies for over-fitting by adding noise during training stage, constructing realistic mock datasets and performing a extensive grid-search over the hyper-parameters of the network.

In the future, we plan to provide thermal parameter constraints using observational spectra at $4\leq z\leq5$. We have shown a glimpse of IGM gas conditions reconstruction from a real spectrum in Section~\ref{sec:obs}.  This unique approach enables insight into the IGM using individual 20 \mpc segments contrary to current methods which require averaging together a much larger volume of the universe.
Another possible implication of our method is to perform reconstruction for models with thermal fluctuations at $z=5.0$.  This method would require model grids of inhomogeneous reionization simulations on much larger scales. Potentially we can see evidence of excess scatter in the distribution of recovered thermal parameters along individual sightlines.


\appendix

\section{Comparison with real-space distributions}

\begin{figure*}
\resizebox{17.6cm}{!}
{\includegraphics[width=\columnwidth,trim={0cm .cm 0cm 0},clip]{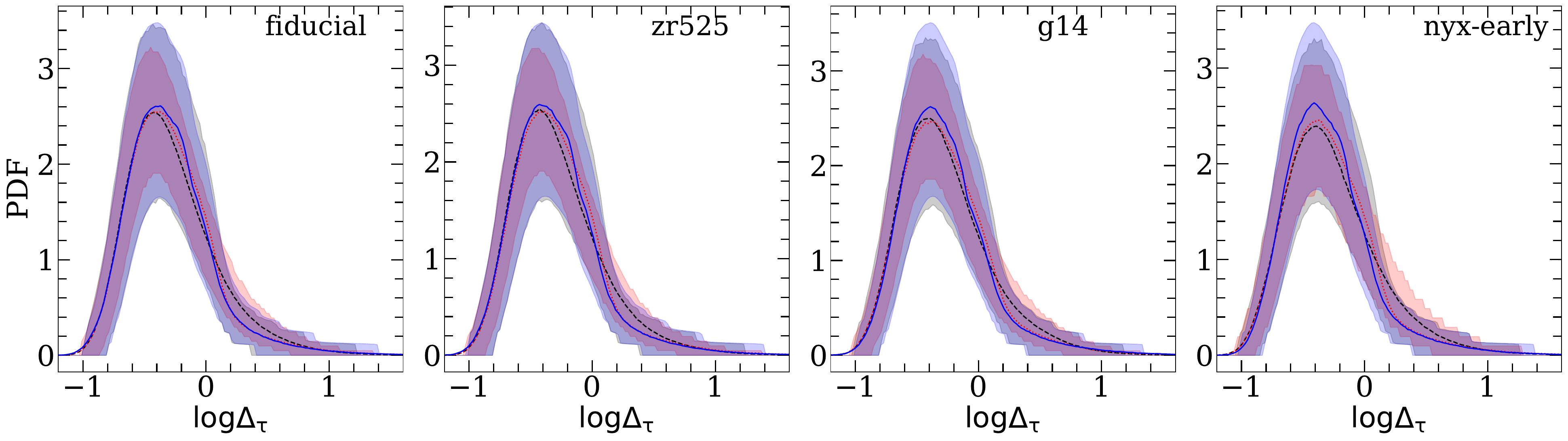}}
\resizebox{17.6cm}{!}
{\includegraphics[width=\columnwidth,trim={0cm .cm 0cm 0},clip]{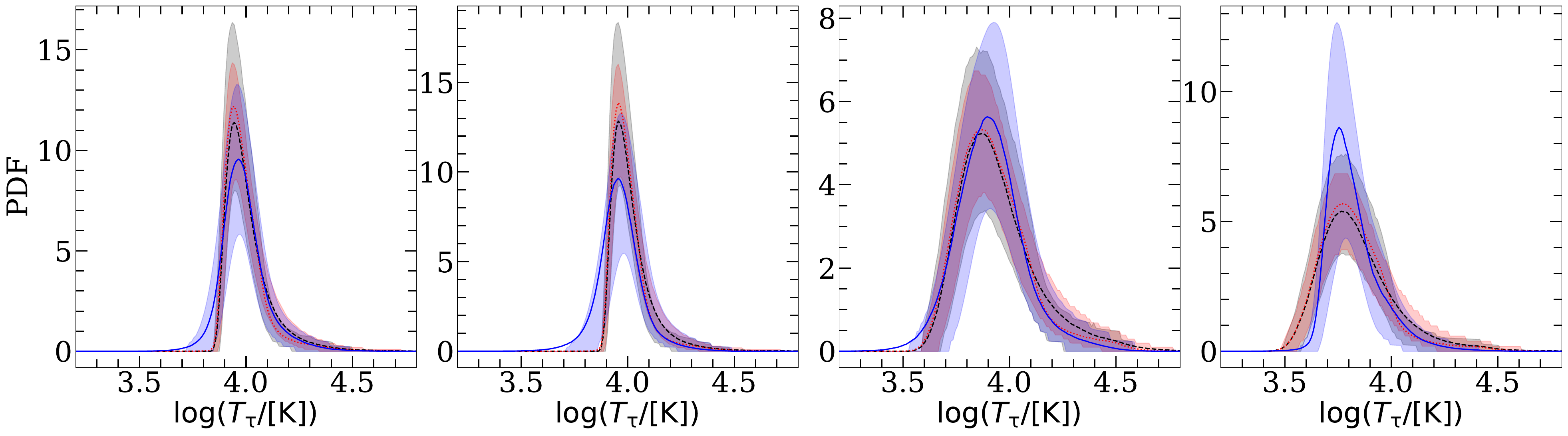}}
\vspace*{-2mm}
\caption{Same as for Figure~\ref{fig:dist}, but now for selected models shown in legends. Each panel shows real-space (red), \lya optical depth-weighted (black) and predicted quantities (blue). }
\label{fig:dist_comparison}
\end{figure*}

\begin{figure*}
\resizebox{17.6cm}{!}
{\includegraphics[width=\columnwidth,trim={0cm 0cm 0cm 0},clip]{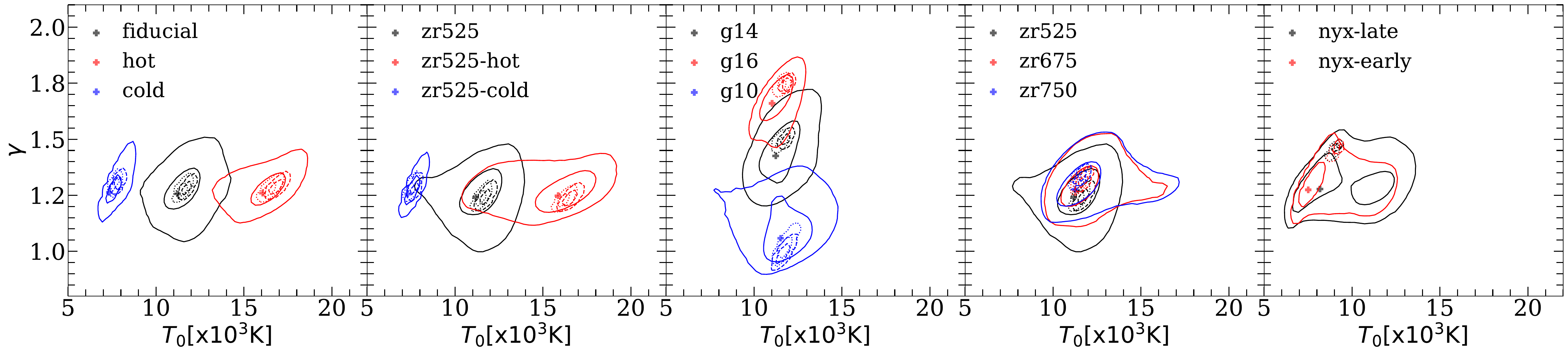}}
\vspace*{-2mm}
\caption{Same as for Figure~\ref{fig:thermal_params} but now a comparison between real-space (dashed), \lya optical depth-weighted (dotted) and predicted quantities (solid) with ${\rm S/N=50}$. The contours enclose 1$\sigma$ and 2$\sigma$ scatter over 20\mpc skewers. }
\label{fig:thermal_params_comparison}
\end{figure*}

In this paper we have chosen to work with optical depth-weighted quantities and have determined thermal parameters using predicted \optd-\optt realizations. In actuality, the weighted quantities are just a proxy for real-space quantities i.e \temp-\den. So, a comparison of density, temperature and \to-$\gamma$ distributions using real-space and optical depth-weighed quantities is presented in this section. We have summarised the results in Figure~\ref{fig:dist_comparison} and Figure~\ref{fig:thermal_params_comparison}.

In Figure~\ref{fig:dist_comparison}, we have compared real-space (red dotted), \lya optical depth-weighted (black dashed) and predicted (blue solid) quantities for density (top row) and temperature (bottom row) distributions. The shaded regions represents the 1$\sigma$ scatter over 20 \mpc skewers, which is appropriately colored. 

As expected, the IGM density distributions (comparing between curves across panels in first row) are very similar. Furthermore, the real-space, optical depth-weighted and predicted for given model (comparing curves in single panel) also closely match. Although, the distributions of optical depth-weighted density (blue curves) have a slight tail at the high-density end, \optd$\simeq0.2$. The reason is, the act of optical depth-weighting shifts slightly under-dense gas into mild over-densities. A comparison of the shaded regions suggests that all quantities exhibit a very similar scatter over 20\mpc skewers.

The real-space (red) and optical depth-weighted (black) temperature (bottom row in Figure~\ref{fig:dist_comparison}) also indicate a very similar trend, where the latter has tail at the higher-temperature end. In addition, there is significantly more scatter at around median temperatures in the optical depth-weighted case for the reasons discussed before. Overall, the predicted distributions are broader than the rest and show significantly more scatter as well.
Note, that at the low-temperature end the distribution cannot capture the sharp rise for models with relatively lower $\gamma$ (panels 1 through 2) which gives a noticeable tail. The reason is models with relatively shallower slope have a small range of temperatures which corresponds to a large range of densities.

Figure~\ref{fig:thermal_params_comparison} shows a comparison of \to-$\gamma$ distributions using \den-\temp (real-space), \optd-\optt (optically weighted) and realizations of \optd-\optt plane. The 68th and 95th percentiles of these distributions are shown as dashed, dotted and solid contours respectively.

It is obvious that the predicted \to-$\gamma$ distributions are broader than the rest. This reassures us that we do not have to incorporate additional uncertainties from our choice of optical depth-weighted quantities into our predictions. Upon close inspection we can see a subtle bias between actual optical depth-weighted and real-space distributions. The former has slightly lower value of \to ($\sim500{\rm K}$), for majority of models. Notice, that this also causes the predicted distributions to slight under-predict \to for majority of models, which can be seen by comparing the plus symbol with the dashed contours. Despite these subtle biases the real-space distribution broadly lies within 1$\sigma$ of the predicted optical depth-weighted distributions.

\section{Mean flux tests}

\begin{figure*}
\resizebox{17.6cm}{!}
{\includegraphics[width=\columnwidth,trim={0cm .cm 0cm 0},clip]{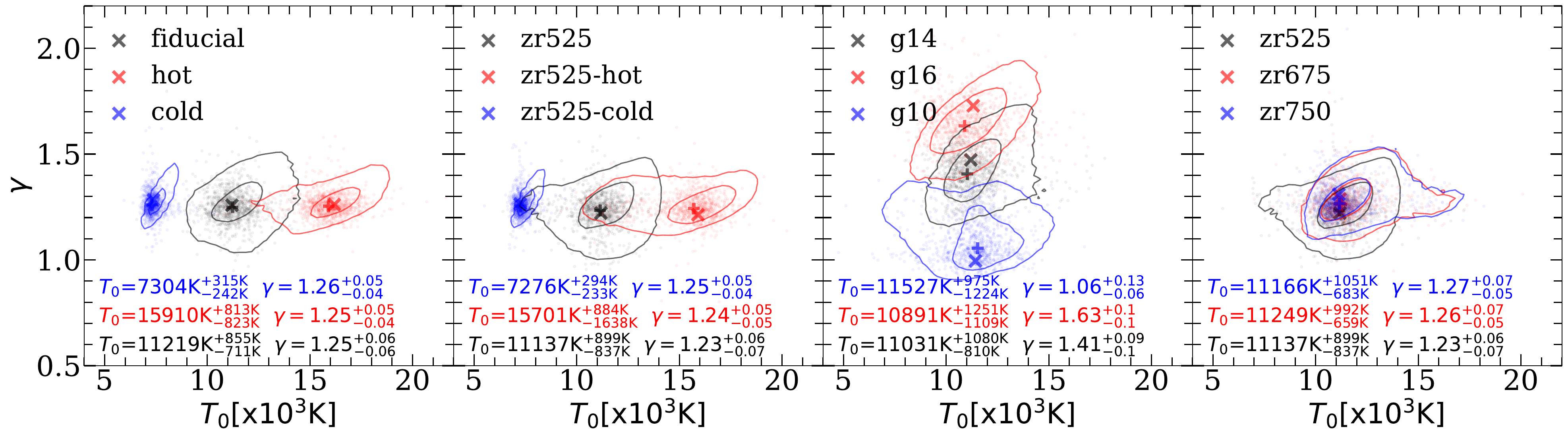}}
\resizebox{17.6cm}{!}
{\includegraphics[width=\columnwidth,trim={0cm .cm 0cm 0},clip]{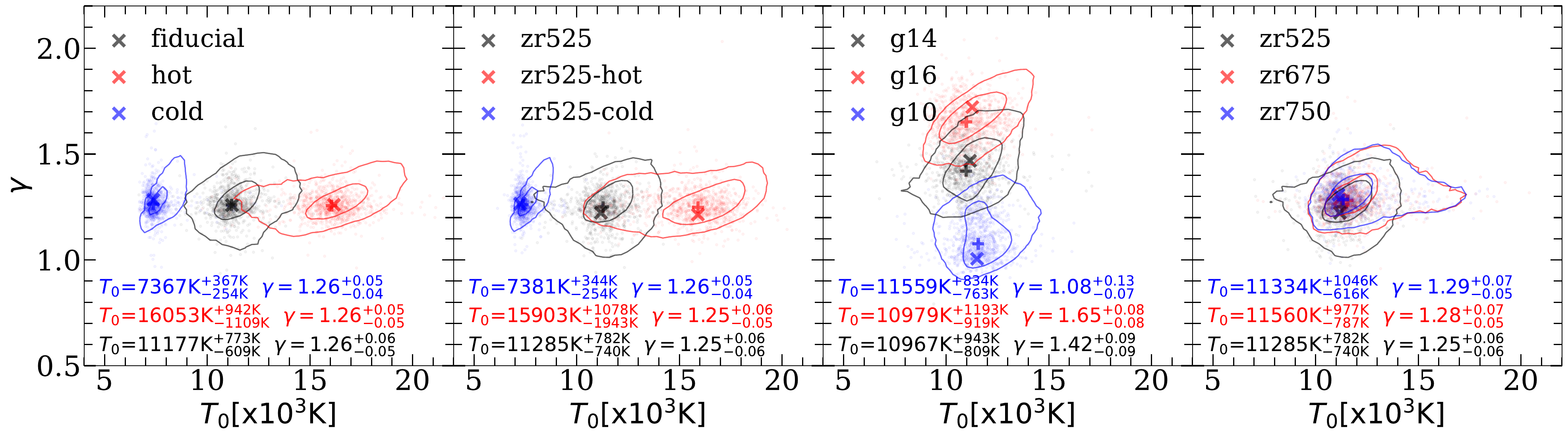}}
\vspace*{-2mm}
\caption{Same as for Figure~\ref{fig:thermal_params}, but now for \lya flux matched to mean flux 10 percent higher (top) and lower (bottom) than fiducial value at $z=4.0$ with ${\rm S/N=50}$ per pixel.}
\label{fig:mean_flux_comparison}
\end{figure*}

To quantity the impact of uncertainties on the mean flux on our \to-$\gamma$ predictions, we rescale \lya flux to match $\langle F \rangle=0.46805$ and $0.38295$, which is 10 percent higher and lower than our value for ${\rm S/N}=50$ per pixel case. These rescaled datasets are used to provide the predictions from network with frozen weights and shown in Figure~\ref{fig:mean_flux_comparison}.
Overall, there is no noticeable, however, there is some subtle change in $\gamma$ which is under-predicted most noticeably (2 to 3 percent) for higher mean flux case.

\section{Box size test}
\begin{figure}
\includegraphics[width=.7\columnwidth,trim={0.0cm 0.0cm 0.0cm 0},clip]{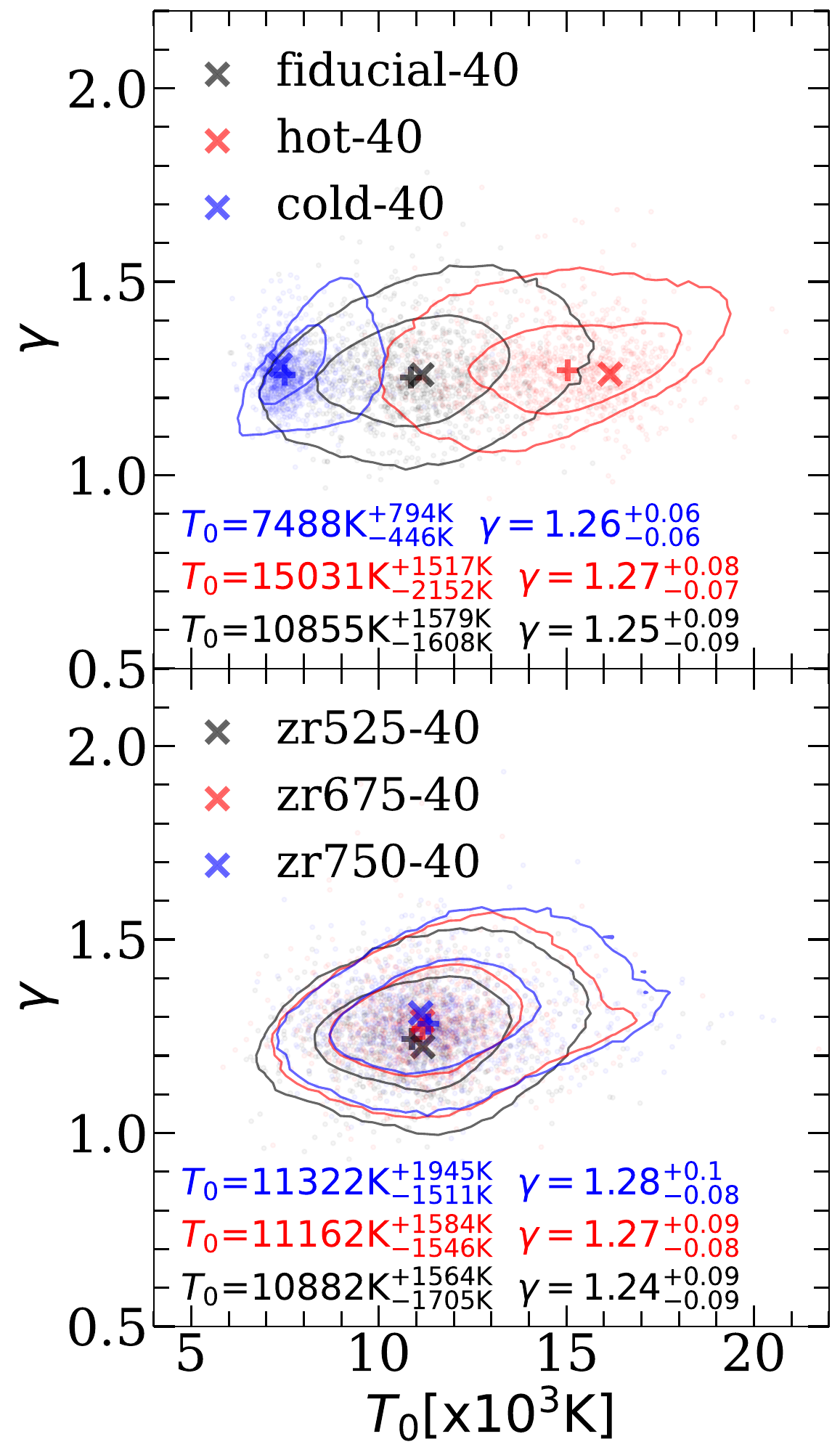}
\vspace*{-2mm}
\caption{Same as for Figure~\ref{fig:thermal_params}, but now for 40\mpc boxes from Sherwood-Relics at $z=4$  with ${\rm S/N=50}$ per pixel.}
\label{fig:box_size}
\end{figure}

We have taken six more runs from Sherwood-Relics to see the impact of box size on the networks predictions with fixed mass resolution. These runs has box length of 40\mpc and has $2048^3$ gas and dark matter particles.
We have taken 20\mpc long skewers from these boxes for this exercise. These skewers are only used at the prediction stage.
Our network relies on the periodicity of skewers during training which is not the case for these skewers taken from 40\mpc box. Therefore, we slightly modified our training by masking (16 pixels) at start of \lya skewer during training with our 20\mpc boxes to break dependence of network on periodicity at boundaries. The masking was done after each skewer is periodically shifted by a random value. 
We use this slightly modified network to obtain predictions for 40\mpc runs (model shown in legend) as shown \to-$\gamma$ distributions in Figure~\ref{fig:box_size}. Although, the mean values remain largely unchanged, the predicted distributions are in fairly broad. This in partly due to network cannot rely on periodic boundary and partly due to box size impacting the \to-$\gamma$ prediction.

\section*{Acknowledgments}
We thank Jose Onorbe for sharing the {\sc nyx} simulations and useful discussion.
The Sherwood-Relics simulations and its post-processing were performed using the Curie supercomputer at the
Tre Grand Centre de Calcul (TGCC), and the DiRAC Data Analytic system at the
University of Cambridge, operated by the University of Cambridge High
Performance Computing Service on behalf of the STFC DiRAC HPC Facility
(www.dirac.ac.uk). This equipment was funded by BIS National E-infrastructure
capital grant (ST/K001590/1), STFC capital grants ST/H008861/1 and
ST/H00887X/1, and STFC DiRAC Operations grant ST/K00333X/1. DiRAC is part of
the National E- Infrastructure.
Computations in this work were also  performed using the CALX machines at 
IoA.  Support by ERC Advanced Grant 320596 `The Emergence of Structure During 
the Epoch of reionization' is gratefully acknowledged. 

\section*{Data Availability}
The data underlying this article will be shared on reasonable request to the corresponding author.

\bibliographystyle{mnras}
\bibliography{bibliography}

\begin{thebibliography}{}
\makeatletter
\relax
\def\mn@urlcharsother{\let\do\@makeother \do\$\do\&\do\#\do\^\do\_\do\%\do\~}
\def\mn@doi{\begingroup\mn@urlcharsother \@ifnextchar [ {\mn@doi@}
  {\mn@doi@[]}}
\def\mn@doi@[#1]#2{\def\@tempa{#1}\ifx\@tempa\@empty \href
  {http://dx.doi.org/#2} {doi:#2}\else \href {http://dx.doi.org/#2} {#1}\fi
  \endgroup}
\def\mn@eprint#1#2{\mn@eprint@#1:#2::\@nil}
\def\mn@eprint@arXiv#1{\href {http://arxiv.org/abs/#1} {{\tt arXiv:#1}}}
\def\mn@eprint@dblp#1{\href {http://dblp.uni-trier.de/rec/bibtex/#1.xml}
  {dblp:#1}}
\def\mn@eprint@#1:#2:#3:#4\@nil{\def\@tempa {#1}\def\@tempb {#2}\def\@tempc
  {#3}\ifx \@tempc \@empty \let \@tempc \@tempb \let \@tempb \@tempa \fi \ifx
  \@tempb \@empty \def\@tempb {arXiv}\fi \@ifundefined
  {mn@eprint@\@tempb}{\@tempb:\@tempc}{\expandafter \expandafter \csname
  mn@eprint@\@tempb\endcsname \expandafter{\@tempc}}}

\bibitem[\protect\citeauthoryear{{Almgren}, {Bell}, {Lijewski}, {Luki{\'c}}  \&
  {Van Andel}}{{Almgren} et~al.}{2013}]{Almgren2013}
{Almgren} A.~S.,  {Bell} J.~B.,  {Lijewski} M.~J.,  {Luki{\'c}} Z.,   {Van
  Andel} E.,  2013, \mn@doi [\apj] {10.1088/0004-637X/765/1/39}, \href
  {https://ui.adsabs.harvard.edu/abs/2013ApJ...765...39A} {765, 39}

\bibitem[\protect\citeauthoryear{{Becker} \& {Bolton}}{{Becker} \&
  {Bolton}}{2013}]{Becker_2013MNRAS}
{Becker} G.~D.,  {Bolton} J.~S.,  2013, \mn@doi [\mnras]
  {10.1093/mnras/stt1610}, \href
  {http://adsabs.harvard.edu/abs/2013MNRAS.436.1023B} {436, 1023}

\bibitem[\protect\citeauthoryear{{Becker}, {Bolton}, {Haehnelt}  \&
  {Sargent}}{{Becker} et~al.}{2011}]{Becker_2011MNRAS}
{Becker} G.~D.,  {Bolton} J.~S.,  {Haehnelt} M.~G.,   {Sargent} W.~L.~W.,
  2011, \mn@doi [\mnras] {10.1111/j.1365-2966.2010.17507.x}, 410, 1096

\bibitem[\protect\citeauthoryear{{Becker}, {Bolton}  \& {Lidz}}{{Becker}
  et~al.}{2015}]{Becker_2015PASA}
{Becker} G.~D.,  {Bolton} J.~S.,   {Lidz} A.,  2015, \mn@doi [PASA]
  {10.1017/pasa.2015.45}, 32, e045

\bibitem[\protect\citeauthoryear{{Boera}, {Murphy}, {Becker}  \&
  {Bolton}}{{Boera} et~al.}{2014}]{Boera_2014MNRAS}
{Boera} E.,  {Murphy} M.~T.,  {Becker} G.~D.,   {Bolton} J.~S.,  2014, \mn@doi
  [\mnras] {10.1093/mnras/stu660}, \href
  {https://ui.adsabs.harvard.edu/abs/2014MNRAS.441.1916B} {441, 1916}

\bibitem[\protect\citeauthoryear{{Boera}, {Murphy}, {Becker}  \&
  {Bolton}}{{Boera} et~al.}{2016}]{Boera_2016MNRAS}
{Boera} E.,  {Murphy} M.~T.,  {Becker} G.~D.,   {Bolton} J.~S.,  2016, \mn@doi
  [\mnras] {10.1093/mnrasl/slv172}, \href
  {http://adsabs.harvard.edu/abs/2016MNRAS.456L..79B} {456, L79}

\bibitem[\protect\citeauthoryear{{Boera}, {Becker}, {Bolton}  \&
  {Nasir}}{{Boera} et~al.}{2019}]{Boera_2019ApJ}
{Boera} E.,  {Becker} G.~D.,  {Bolton} J.~S.,   {Nasir} F.,  2019, \mn@doi
  [\apj] {10.3847/1538-4357/aafee4}, \href
  {https://ui.adsabs.harvard.edu/abs/2019ApJ...872..101B} {872, 101}

\bibitem[\protect\citeauthoryear{{Bolton}, {Viel}, {Kim}, {Haehnelt}  \&
  {Carswell}}{{Bolton} et~al.}{2008}]{Bolton2008}
{Bolton} J.~S.,  {Viel} M.,  {Kim} T.~S.,  {Haehnelt} M.~G.,   {Carswell}
  R.~F.,  2008, \mn@doi [\mnras] {10.1111/j.1365-2966.2008.13114.x}, \href
  {https://ui.adsabs.harvard.edu/abs/2008MNRAS.386.1131B} {386, 1131}

\bibitem[\protect\citeauthoryear{{Bolton}, {Becker}, {Raskutti}, {Wyithe},
  {Haehnelt}  \& {Sargent}}{{Bolton} et~al.}{2012}]{Bolton_2012MNRAS}
{Bolton} J.~S.,  {Becker} G.~D.,  {Raskutti} S.,  {Wyithe} J.~S.~B.,
  {Haehnelt} M.~G.,   {Sargent} W.~L.~W.,  2012, \mn@doi [\mnras]
  {10.1111/j.1365-2966.2011.19929.x}, \href
  {http://adsabs.harvard.edu/abs/2012MNRAS.419.2880B} {419, 2880}

\bibitem[\protect\citeauthoryear{{Bolton}, {Becker}, {Haehnelt}  \&
  {Viel}}{{Bolton} et~al.}{2014}]{Bolton_2014MNRAS}
{Bolton} J.~S.,  {Becker} G.~D.,  {Haehnelt} M.~G.,   {Viel} M.,  2014, \mn@doi
  [\mnras] {10.1093/mnras/stt2374}, 438, 2499

\bibitem[\protect\citeauthoryear{{Bolton}, {Puchwein}, {Sijacki}, {Haehnelt},
  {Kim}, {Meiksin}, {Regan}  \& {Viel}}{{Bolton}
  et~al.}{2017}]{Bolton_2017MNRAS}
{Bolton} J.~S.,  {Puchwein} E.,  {Sijacki} D.,  {Haehnelt} M.~G.,  {Kim} T.-S.,
   {Meiksin} A.,  {Regan} J.~A.,   {Viel} M.,  2017, \mn@doi [\mnras]
  {10.1093/mnras/stw2397}, 464, 897

\bibitem[\protect\citeauthoryear{{Bosman}, {Fan}, {Jiang}, {Reed}, {Matsuoka},
  {Becker}  \& {Haehnelt}}{{Bosman} et~al.}{2018}]{Bosman_2018MNRAS}
{Bosman} S. E.~I.,  {Fan} X.,  {Jiang} L.,  {Reed} S.,  {Matsuoka} Y.,
  {Becker} G.,   {Haehnelt} M.,  2018, \mn@doi [\mnras]
  {10.1093/mnras/sty1344}, 479, 1055

\bibitem[\protect\citeauthoryear{{Calura}, {Tescari}, {D'Odorico}, {Viel},
  {Cristiani}, {Kim}  \& {Bolton}}{{Calura} et~al.}{2012}]{Calura2012}
{Calura} F.,  {Tescari} E.,  {D'Odorico} V.,  {Viel} M.,  {Cristiani} S.,
  {Kim} T.-S.,   {Bolton} J.~S.,  2012, \mn@doi [\mnras]
  {10.1111/j.1365-2966.2012.20811.x}, 422, 3019

\bibitem[\protect\citeauthoryear{Chollet et~al.}{Chollet
  et~al.}{2015}]{chollet2015keras}
Chollet F.,  et~al., 2015, Keras, \url {https://github.com/fchollet/keras}

\bibitem[\protect\citeauthoryear{{Croft}, {Weinberg}, {Bolte}, {Burles},
  {Hernquist}, {Katz}, {Kirkman}  \& {Tytler}}{{Croft}
  et~al.}{2002}]{Croft_2002ApJ}
{Croft} R.~A.~C.,  {Weinberg} D.~H.,  {Bolte} M.,  {Burles} S.,  {Hernquist}
  L.,  {Katz} N.,  {Kirkman} D.,   {Tytler} D.,  2002, \mn@doi [\apj]
  {10.1086/344099}, 581, 20

\bibitem[\protect\citeauthoryear{{D'Aloisio}, {McQuinn}  \& {Trac}}{{D'Aloisio}
  et~al.}{2015}]{D'Aloisio2015}
{D'Aloisio} A.,  {McQuinn} M.,   {Trac} H.,  2015, \mn@doi [\apjl]
  {10.1088/2041-8205/813/2/L38}, \href
  {https://ui.adsabs.harvard.edu/abs/2015ApJ...813L..38D} {813, L38}

\bibitem[\protect\citeauthoryear{{Eilers}, {Hogg}, {Sch{\"o}lkopf},
  {Foreman-Mackey}, {Davies}  \& {Schindler}}{{Eilers}
  et~al.}{2022}]{Eilers2022}
{Eilers} A.-C.,  {Hogg} D.~W.,  {Sch{\"o}lkopf} B.,  {Foreman-Mackey} D.,
  {Davies} F.~B.,   {Schindler} J.-T.,  2022, \mn@doi [\apj]
  {10.3847/1538-4357/ac8ead}, \href
  {https://ui.adsabs.harvard.edu/abs/2022ApJ...938...17E} {938, 17}

\bibitem[\protect\citeauthoryear{{Gaikwad}, {Srianand}, {Choudhury}  \&
  {Khaire}}{{Gaikwad} et~al.}{2017}]{Gaikwad17}
{Gaikwad} P.,  {Srianand} R.,  {Choudhury} T.~R.,   {Khaire} V.,  2017, \mn@doi
  [\mnras] {10.1093/mnras/stx248}, 467, 3172

\bibitem[\protect\citeauthoryear{{Gaikwad} et~al.,}{{Gaikwad}
  et~al.}{2020}]{Gaikwad_2020}
{Gaikwad} P.,  et~al., 2020, \mn@doi [\mnras] {10.1093/mnras/staa907}, \href
  {https://ui.adsabs.harvard.edu/abs/2020MNRAS.494.5091G} {494, 5091}

\bibitem[\protect\citeauthoryear{{Gaikwad}, {Srianand}, {Haehnelt}  \&
  {Choudhury}}{{Gaikwad} et~al.}{2021}]{Gaikwad_2021}
{Gaikwad} P.,  {Srianand} R.,  {Haehnelt} M.~G.,   {Choudhury} T.~R.,  2021,
  \mn@doi [\mnras] {10.1093/mnras/stab2017}, \href
  {https://ui.adsabs.harvard.edu/abs/2021MNRAS.506.4389G} {506, 4389}

\bibitem[\protect\citeauthoryear{{Garzilli}, {Bolton}, {Kim}, {Leach}  \&
  {Viel}}{{Garzilli} et~al.}{2012}]{Garzilli_2012MNRAS}
{Garzilli} A.,  {Bolton} J.~S.,  {Kim} T.-S.,  {Leach} S.,   {Viel} M.,  2012,
  \mn@doi [\mnras] {10.1111/j.1365-2966.2012.21223.x}, 424, 1723

\bibitem[\protect\citeauthoryear{Goodfellow, Bengio  \& Courville}{Goodfellow
  et~al.}{2016}]{Goodfellow2016}
Goodfellow I.~J.,  Bengio Y.,   Courville A.,  2016, Deep Learning.
MIT Press, Cambridge, MA, USA

\bibitem[\protect\citeauthoryear{{Haardt} \& {Madau}}{{Haardt} \&
  {Madau}}{2012}]{Haardt2012}
{Haardt} F.,  {Madau} P.,  2012, \mn@doi [\apj] {10.1088/0004-637X/746/2/125},
  746, 125

\bibitem[\protect\citeauthoryear{{Haehnelt} \& {Steinmetz}}{{Haehnelt} \&
  {Steinmetz}}{1998}]{Haehnelt_1998MNRAS}
{Haehnelt} M.~G.,  {Steinmetz} M.,  1998, \mn@doi [\mnras]
  {10.1046/j.1365-8711.1998.01879.x}, 298, L21

\bibitem[\protect\citeauthoryear{{Harrington}, {Mustafa}, {Dornfest},
  {Horowitz}  \& {Luki{\'c}}}{{Harrington} et~al.}{2022}]{Harrington2022}
{Harrington} P.,  {Mustafa} M.,  {Dornfest} M.,  {Horowitz} B.,   {Luki{\'c}}
  Z.,  2022, \mn@doi [\apj] {10.3847/1538-4357/ac5faa}, \href
  {https://ui.adsabs.harvard.edu/abs/2022ApJ...929..160H} {929, 160}

\bibitem[\protect\citeauthoryear{{He}, {Zhang}, {Ren}  \& {Sun}}{{He}
  et~al.}{2015}]{He2015arXiv}
{He} K.,  {Zhang} X.,  {Ren} S.,   {Sun} J.,  2015, \mn@doi [arXiv e-prints]
  {10.48550/arXiv.1512.03385}, \href
  {https://ui.adsabs.harvard.edu/abs/2015arXiv151203385H} {p. arXiv:1512.03385}

\bibitem[\protect\citeauthoryear{{Hiss}, {Walther}, {Hennawi}, {O{\~n}orbe},
  {O'Meara}, {Rorai}  \& {Luki{\'c}}}{{Hiss} et~al.}{2018}]{Hiss2018}
{Hiss} H.,  {Walther} M.,  {Hennawi} J.~F.,  {O{\~n}orbe} J.,  {O'Meara} J.~M.,
   {Rorai} A.,   {Luki{\'c}} Z.,  2018, \mn@doi [\apj]
  {10.3847/1538-4357/aada86}, \href
  {https://ui.adsabs.harvard.edu/abs/2018ApJ...865...42H} {865, 42}

\bibitem[\protect\citeauthoryear{{Hiss}, {Walther}, {O{\~n}orbe}  \&
  {Hennawi}}{{Hiss} et~al.}{2019}]{Hiss2019}
{Hiss} H.,  {Walther} M.,  {O{\~n}orbe} J.,   {Hennawi} J.~F.,  2019, \mn@doi
  [\apj] {10.3847/1538-4357/ab1418}, \href
  {https://ui.adsabs.harvard.edu/abs/2019ApJ...876...71H} {876, 71}

\bibitem[\protect\citeauthoryear{{Huang}, {Croft}  \& {Arora}}{{Huang}
  et~al.}{2021}]{Huang2021}
{Huang} L.,  {Croft} R. A.~C.,   {Arora} H.,  2021, \mn@doi [\mnras]
  {10.1093/mnras/stab2041}, \href
  {https://ui.adsabs.harvard.edu/abs/2021MNRAS.506.5212H} {506, 5212}

\bibitem[\protect\citeauthoryear{{Hui} \& {Gnedin}}{{Hui} \&
  {Gnedin}}{1997}]{Hui_1997MNRAS}
{Hui} L.,  {Gnedin} N.~Y.,  1997, \mnras, 292, 27

\bibitem[\protect\citeauthoryear{{Ir{\v{s}}i{\v{c}}}
  et~al.,}{{Ir{\v{s}}i{\v{c}}} et~al.}{2017}]{Irsic2017}
{Ir{\v{s}}i{\v{c}}} V.,  et~al., 2017, \mn@doi [\prd]
  {10.1103/PhysRevD.96.023522}, \href
  {https://ui.adsabs.harvard.edu/abs/2017PhRvD..96b3522I} {96, 023522}

\bibitem[\protect\citeauthoryear{{Ir{\v{s}}i{\v{c}}}
  et~al.,}{{Ir{\v{s}}i{\v{c}}} et~al.}{2023}]{Irsic2023}
{Ir{\v{s}}i{\v{c}}} V.,  et~al., 2023, \mn@doi [arXiv e-prints]
  {10.48550/arXiv.2309.04533}, \href
  {https://ui.adsabs.harvard.edu/abs/2023arXiv230904533I} {p. arXiv:2309.04533}

\bibitem[\protect\citeauthoryear{{Keating}, {Puchwein}  \&
  {Haehnelt}}{{Keating} et~al.}{2018}]{Keating2018}
{Keating} L.~C.,  {Puchwein} E.,   {Haehnelt} M.~G.,  2018, \mn@doi [\mnras]
  {10.1093/mnras/sty968}, \href
  {https://ui.adsabs.harvard.edu/abs/2018MNRAS.477.5501K} {477, 5501}

\bibitem[\protect\citeauthoryear{{Lee} et~al.,}{{Lee}
  et~al.}{2015}]{Lee_2015ApJ}
{Lee} K.-G.,  et~al., 2015, \mn@doi [\apj] {10.1088/0004-637X/799/2/196}, 799,
  196

\bibitem[\protect\citeauthoryear{{Lidz}, {Heitmann}, {Hui}, {Habib}, {Rauch}
  \& {Sargent}}{{Lidz} et~al.}{2006}]{Lidz_2006ApJ}
{Lidz} A.,  {Heitmann} K.,  {Hui} L.,  {Habib} S.,  {Rauch} M.,   {Sargent}
  W.~L.~W.,  2006, \mn@doi [\apj] {10.1086/498699}, \href
  {http://adsabs.harvard.edu/abs/2006ApJ...638...27L} {638, 27}

\bibitem[\protect\citeauthoryear{{Lidz}, {Faucher-Gigu{\`e}re}, {Dall'Aglio},
  {McQuinn}, {Fechner}, {Zaldarriaga}, {Hernquist}  \& {Dutta}}{{Lidz}
  et~al.}{2010}]{Lidz_2010ApJ}
{Lidz} A.,  {Faucher-Gigu{\`e}re} C.-A.,  {Dall'Aglio} A.,  {McQuinn} M.,
  {Fechner} C.,  {Zaldarriaga} M.,  {Hernquist} L.,   {Dutta} S.,  2010,
  \mn@doi [\apj] {10.1088/0004-637X/718/1/199}, 718, 199

\bibitem[\protect\citeauthoryear{{McDonald}, {Miralda-Escud{\'e}}, {Rauch},
  {Sargent}, {Barlow}  \& {Cen}}{{McDonald} et~al.}{2001}]{Mcdonald_2001ApJ}
{McDonald} P.,  {Miralda-Escud{\'e}} J.,  {Rauch} M.,  {Sargent} W.~L.~W.,
  {Barlow} T.~A.,   {Cen} R.,  2001, \mn@doi [\apj] {10.1086/323426}, 562, 52

\bibitem[\protect\citeauthoryear{{McQuinn} \& {Upton Sanderbeck}}{{McQuinn} \&
  {Upton Sanderbeck}}{2016}]{McQuinnSanderbeck2016}
{McQuinn} M.,  {Upton Sanderbeck} P.~R.,  2016, \mn@doi [\mnras]
  {10.1093/mnras/stv2675}, 456, 47

\bibitem[\protect\citeauthoryear{{Meiksin}}{{Meiksin}}{2000}]{Meiksin2000}
{Meiksin} A.,  2000, \mn@doi [\mnras] {10.1046/j.1365-8711.2000.03315.x}, 314,
  566

\bibitem[\protect\citeauthoryear{{Miralda-Escud{\'e}} \&
  {Rees}}{{Miralda-Escud{\'e}} \& {Rees}}{1994}]{Miralda-Escude1994}
{Miralda-Escud{\'e}} J.,  {Rees} M.~J.,  1994, \mn@doi [\mnras]
  {10.1093/mnras/266.2.343}, \href
  {https://ui.adsabs.harvard.edu/abs/1994MNRAS.266..343M} {266, 343}

\bibitem[\protect\citeauthoryear{{Murphy}, {Kacprzak}, {Savorgnan}  \&
  {Carswell}}{{Murphy} et~al.}{2019}]{Murphy2019}
{Murphy} M.~T.,  {Kacprzak} G.~G.,  {Savorgnan} G. A.~D.,   {Carswell} R.~F.,
  2019, \mn@doi [\mnras] {10.1093/mnras/sty2834}, \href
  {https://ui.adsabs.harvard.edu/abs/2019MNRAS.482.3458M} {482, 3458}

\bibitem[\protect\citeauthoryear{{Nasir}, {Bolton}  \& {Becker}}{{Nasir}
  et~al.}{2016}]{Nasir_2016MNRAS}
{Nasir} F.,  {Bolton} J.~S.,   {Becker} G.~D.,  2016, \mn@doi [\mnras]
  {10.1093/mnras/stw2147}, 463, 2335

\bibitem[\protect\citeauthoryear{{Nayak}, {Walther}, {Gruen}  \&
  {Adiraju}}{{Nayak} et~al.}{2023}]{Parth2023arXiv}
{Nayak} P.,  {Walther} M.,  {Gruen} D.,   {Adiraju} S.,  2023, \mn@doi [arXiv
  e-prints] {10.48550/arXiv.2311.02167}, \href
  {https://ui.adsabs.harvard.edu/abs/2023arXiv231102167N} {p. arXiv:2311.02167}

\bibitem[\protect\citeauthoryear{{O{\~n}orbe}, {Hennawi}  \&
  {Luki{\'c}}}{{O{\~n}orbe} et~al.}{2017}]{Onorbe2017}
{O{\~n}orbe} J.,  {Hennawi} J.~F.,   {Luki{\'c}} Z.,  2017, \mn@doi [\apj]
  {10.3847/1538-4357/aa6031}, \href
  {https://ui.adsabs.harvard.edu/abs/2017ApJ...837..106O} {837, 106}

\bibitem[\protect\citeauthoryear{{Padmanabhan}, {Srianand}  \&
  {Choudhury}}{{Padmanabhan} et~al.}{2015}]{Padmanabhan2015}
{Padmanabhan} H.,  {Srianand} R.,   {Choudhury} T.~R.,  2015, \mn@doi [\mnras]
  {10.1093/mnrasl/slv041}, \href
  {https://ui.adsabs.harvard.edu/abs/2015MNRAS.450L..29P} {450, L29}

\bibitem[\protect\citeauthoryear{{Peeples}, {Weinberg}, {Dav{\'e}}, {Fardal}
  \& {Katz}}{{Peeples} et~al.}{2010}]{Peeples_2010MNRAS}
{Peeples} M.~S.,  {Weinberg} D.~H.,  {Dav{\'e}} R.,  {Fardal} M.~A.,   {Katz}
  N.,  2010, \mn@doi [\mnras] {10.1111/j.1365-2966.2010.16383.x}, 404, 1281

\bibitem[\protect\citeauthoryear{{Puchwein}, {Bolton}, {Haehnelt}, {Madau},
  {Becker}  \& {Haardt}}{{Puchwein} et~al.}{2015}]{Puchwein_2015MNRAS}
{Puchwein} E.,  {Bolton} J.~S.,  {Haehnelt} M.~G.,  {Madau} P.,  {Becker}
  G.~D.,   {Haardt} F.,  2015, \mn@doi [\mnras] {10.1093/mnras/stv773}, 450,
  4081

\bibitem[\protect\citeauthoryear{{Puchwein}, {Haardt}, {Haehnelt}  \&
  {Madau}}{{Puchwein} et~al.}{2019}]{Puchwein2019}
{Puchwein} E.,  {Haardt} F.,  {Haehnelt} M.~G.,   {Madau} P.,  2019, \mn@doi
  [\mnras] {10.1093/mnras/stz222}, \href
  {https://ui.adsabs.harvard.edu/abs/2019MNRAS.485...47P} {485, 47}

\bibitem[\protect\citeauthoryear{{Puchwein} et~al.,}{{Puchwein}
  et~al.}{2023}]{Puchwein2023}
{Puchwein} E.,  et~al., 2023, \mn@doi [\mnras] {10.1093/mnras/stac3761}, \href
  {https://ui.adsabs.harvard.edu/abs/2023MNRAS.519.6162P} {519, 6162}

\bibitem[\protect\citeauthoryear{{Ricotti}, {Gnedin}  \& {Shull}}{{Ricotti}
  et~al.}{2000}]{Ricotti2000}
{Ricotti} M.,  {Gnedin} N.~Y.,   {Shull} J.~M.,  2000, \mn@doi [\apj]
  {10.1086/308733}, \href
  {https://ui.adsabs.harvard.edu/abs/2000ApJ...534...41R} {534, 41}

\bibitem[\protect\citeauthoryear{{Rudie}, {Steidel}  \& {Pettini}}{{Rudie}
  et~al.}{2012}]{Rudie_2012ApJ}
{Rudie} G.~C.,  {Steidel} C.~C.,   {Pettini} M.,  2012, \mn@doi [\apjl]
  {10.1088/2041-8205/757/2/L30}, 757, L30

\bibitem[\protect\citeauthoryear{{Schaye}, {Theuns}, {Rauch}, {Efstathiou}  \&
  {Sargent}}{{Schaye} et~al.}{2000}]{Schaye_2000MNRAS}
{Schaye} J.,  {Theuns} T.,  {Rauch} M.,  {Efstathiou} G.,   {Sargent} W.~L.~W.,
   2000, \mn@doi [\mnras] {10.1046/j.1365-8711.2000.03815.x}, 318, 817

\bibitem[\protect\citeauthoryear{{Springel}}{{Springel}}{2005}]{Springel_2005MNRAS}
{Springel} V.,  2005, \mn@doi [\mnras] {10.1111/j.1365-2966.2005.09655.x}, 364,
  1105

\bibitem[\protect\citeauthoryear{{Telikova}, {Shternin}  \&
  {Balashev}}{{Telikova} et~al.}{2019}]{Telikova2019}
{Telikova} K.~N.,  {Shternin} P.~S.,   {Balashev} S.~A.,  2019, \mn@doi [\apj]
  {10.3847/1538-4357/ab52fe}, \href
  {https://ui.adsabs.harvard.edu/abs/2019ApJ...887..205T} {887, 205}

\bibitem[\protect\citeauthoryear{{Theuns}, {Schaye}  \& {Haehnelt}}{{Theuns}
  et~al.}{2000}]{Theuns_2000MNRAS}
{Theuns} T.,  {Schaye} J.,   {Haehnelt} M.~G.,  2000, \mn@doi [\mnras]
  {10.1046/j.1365-8711.2000.03423.x}, 315, 600

\bibitem[\protect\citeauthoryear{{Viel}, {Becker}, {Bolton}  \&
  {Haehnelt}}{{Viel} et~al.}{2013}]{Viel_2013PRD}
{Viel} M.,  {Becker} G.~D.,  {Bolton} J.~S.,   {Haehnelt} M.~G.,  2013, \mn@doi
  [\prd] {10.1103/PhysRevD.88.043502}, 88, 043502

\bibitem[\protect\citeauthoryear{{Villasenor}, {Robertson}, {Madau}  \&
  {Schneider}}{{Villasenor} et~al.}{2023}]{Villasenor2023}
{Villasenor} B.,  {Robertson} B.,  {Madau} P.,   {Schneider} E.,  2023, \mn@doi
  [\prd] {10.1103/PhysRevD.108.023502}, \href
  {https://ui.adsabs.harvard.edu/abs/2023PhRvD.108b3502V} {108, 023502}

\bibitem[\protect\citeauthoryear{{Walther}, {O{\~n}orbe}, {Hennawi}  \&
  {Luki{\'c}}}{{Walther} et~al.}{2019a}]{Walther_2019ApJ}
{Walther} M.,  {O{\~n}orbe} J.,  {Hennawi} J.~F.,   {Luki{\'c}} Z.,  2019a,
  \mn@doi [\apj] {10.3847/1538-4357/aafad1}, \href
  {https://ui.adsabs.harvard.edu/abs/2019ApJ...872...13W} {872, 13}

\bibitem[\protect\citeauthoryear{{Walther}, {O{\~n}orbe}, {Hennawi}  \&
  {Luki{\'c}}}{{Walther} et~al.}{2019b}]{Walther2019}
{Walther} M.,  {O{\~n}orbe} J.,  {Hennawi} J.~F.,   {Luki{\'c}} Z.,  2019b,
  \mn@doi [\apj] {10.3847/1538-4357/aafad1}, \href
  {https://ui.adsabs.harvard.edu/abs/2019ApJ...872...13W} {872, 13}

\bibitem[\protect\citeauthoryear{{Wang}, {Croft}  \& {Shaw}}{{Wang}
  et~al.}{2022}]{Wang2022}
{Wang} R.,  {Croft} R. A.~C.,   {Shaw} P.,  2022, \mn@doi [\mnras]
  {10.1093/mnras/stac1786}, \href
  {https://ui.adsabs.harvard.edu/abs/2022MNRAS.515.1568W} {515, 1568}

\bibitem[\protect\citeauthoryear{{Wolfson}, {Hennawi}, {Davies}, {O{\~n}orbe},
  {Hiss}  \& {Luki{\'c}}}{{Wolfson} et~al.}{2021}]{Wolfson2021}
{Wolfson} M.,  {Hennawi} J.~F.,  {Davies} F.~B.,  {O{\~n}orbe} J.,  {Hiss} H.,
   {Luki{\'c}} Z.,  2021, \mn@doi [\mnras] {10.1093/mnras/stab2920}, \href
  {https://ui.adsabs.harvard.edu/abs/2021MNRAS.508.5493W} {508, 5493}

\bibitem[\protect\citeauthoryear{{Zaldarriaga}}{{Zaldarriaga}}{2002}]{Zaldarriaga2002}
{Zaldarriaga} M.,  2002, \mn@doi [\apj] {10.1086/324212}, 564, 153

\bibitem[\protect\citeauthoryear{{Zaldarriaga}, {Hui}  \&
  {Tegmark}}{{Zaldarriaga} et~al.}{2001}]{Zaldarriaga_2001ApJ}
{Zaldarriaga} M.,  {Hui} L.,   {Tegmark} M.,  2001, \mn@doi [\apj]
  {10.1086/321652}, 557, 519

\bibitem[\protect\citeauthoryear{{Zaroubi}, {Viel}, {Nusser}, {Haehnelt}  \&
  {Kim}}{{Zaroubi} et~al.}{2006}]{Zaroubi_2006MNRAS}
{Zaroubi} S.,  {Viel} M.,  {Nusser} A.,  {Haehnelt} M.,   {Kim} T.-S.,  2006,
  \mn@doi [\mnras] {10.1111/j.1365-2966.2006.10333.x}, \href
  {http://adsabs.harvard.edu/abs/2006MNRAS.369..734Z} {369, 734}

\makeatother
\end{thebibliography}

\appendix

\end{document}